\documentclass[useAMS,usenatbib]{mn2e}
\usepackage{mn2e-breakabs}
\usepackage{graphics}
\usepackage{times}
\newcommand{\hompc}{\,h\,{\rm Mpc}^{-1}}
\newcommand{\mpcoh}{\,h^{-1}\,{\rm Mpc}}
\newcommand{\Hunit}{\,{\rm km}\,{\rm s}^{-1}\,{\rm Mpc}^{-1}}
\newcommand{\simlt}%
{\,\hbox{\lower0.6ex\hbox{$\sim$}\llap{\raise0.6ex\hbox{$<$}}}\,}

\begin{document}

\title[BAO in SDSS DR7]
{Baryon Acoustic Oscillations in the Sloan Digital Sky Survey Data Release 7 Galaxy Sample}

\author[W.J. Percival et al.]{
  \parbox{\textwidth}{
    Will J.\ Percival$^{1}$\thanks{E-mail: will.percival@port.ac.uk},
    Beth A.\ Reid$^{2,3}$,
    Daniel J.\ Eisenstein$^{4}$,
    Neta A.\ Bahcall$^{3}$,
    Tamas Budavari$^{5}$,
    Joshua A.\ Frieman$^{6,7}$,
    Masataka Fukugita$^{8}$,
    James E. Gunn$^{3}$,
    \v{Z}eljko Ivezi\'{c}$^{9}$,
    Gillian R. Knapp$^{3}$,
    Richard G.\ Kron$^{10}$,
    Jon Loveday$^{11}$,
    Robert H.\ Lupton$^{3}$,
    Timothy A.\ McKay$^{12}$,
    Avery Meiksin$^{13}$,
    Robert C.\ Nichol$^{1}$,
    Adrian C.\ Pope$^{14}$,
    David J.\ Schlegel$^{15}$,
    Donald P.\ Schneider$^{16}$,
    David N.\ Spergel$^{3,17}$,
    Chris Stoughton$^{18}$,
    Michael A.\ Strauss$^{3}$,
    Alexander S.\ Szalay$^{5}$,
    Max Tegmark$^{19}$,
    Michael S.\ Vogeley$^{20}$,
    David H.\ Weinberg$^{21}$,
    Donald G.\ York$^{10,22}$,
    Idit Zehavi$^{23}$,
  }
  \vspace*{4pt} \\
  $^{1}$ Institute of Cosmology and Gravitation, University of
  Portsmouth, Dennis Sciama building, Portsmouth, P01 3FX, UK\\
  $^{2}$ Institute of Space Sciences (CSIC-IEEC), UAB, 
  Barcelona 08193, Spain and\\
  Institute for Sciences of the Cosmos (ICC), University of 
  Barcelona, Barcelona 08028, Spain \\
  $^{3}$ Department of Astrophysical Sciences, Princeton University, 
  Princeton, NJ 08544, USA \\
  $^{4}$ Steward Observatory, University of Arizona, 933
  N. Cherry Ave., Tucson, AZ 85121, USA\\
  $^{5}$ Department of Physics and Astronomy, The Johns Hopkins
  University, 3701 San Martin Drive, Baltimore, MD 21218, USA\\
  $^{6}$ Particle Astrophysics Center, Fermilab, P.O. Box 500,
  Batavia, IL 60510, USA\\
  $^{7}$ Kavli Institute for Cosmological Physics, Department of
  Astronomy \& Astrophysics, University of Chicago, Chicago, IL 60637,
  USA\\
  $^{8}$Institute for Cosmic Ray Research, University of Tokyo, 
  Kashiwa 277-8582, Japan\\
  $^{9}$Department of Astronomy, University of Washington
  Box 351580, Seattle, WA 98195, USA\\
  $^{10}$Department of Astronomy and Astrophysics, The University 
  of Chicago, 5640 South Ellis Avenue, Chicago, IL 60637, USA\\
  $^{11}$Astronomy Centre, University of Sussex, Falmer, Brighton, 
  BN1 9QH, UK\\
  $^{12}$Departments of Physics and Astronomy, University of Michigan, 
  Ann Arbor, MI, 48109, USA\\
  $^{13}$ SUPA; Institute for Astronomy, University of Edinburgh,
  Royal Observatory, Blackford Hill, Edinburgh, EH9 3HJ, UK\\
  $^{14}$Los Alamos National Laboratory, PO Box 1663, Los Alamos, 
  NM 87545, USA\\
  $^{15}$Lawrence Berkeley National Lab, 1 Cyclotron Road, 
  MS 50R5032, Berkeley, CA94720, USA\\
  $^{16}$Department of Astronomy and Astrophysics, The
  Pennsylvania State University, University Park, PA 16802, USA\\
  $^{17}$ Princeton Center for Theoretical Science, Princeton
  University, Jadwin Hall, Princeton, NJ 08542 USA\\
  $^{18}$Fermilab, PO Box 500, Batavia, IL  60510, USA\\
  $^{19}$Department of Physics, Massachusetts Institute of
  Technology, Cambridge, MA 02139, USA\\
  $^{20}$ Department of Physics, Drexel University, 
  Philadelphia, PA 19104, USA\\
  $^{21}$Department of Astronomy, The Ohio State University,
  Columbus, OH 43210, USA\\
  $^{22}$Enrico Fermi Institute, University of Chicago, 
  Chicago, IL 60637, USA\\
  $^{23}$ Department of Astronomy, Case Western Reserve University,
  Cleveland, OH 44106, USA}

\date{\today} 

\maketitle

\begin{abstract}

  The spectroscopic Sloan Digital Sky Survey (SDSS) Data Release 7
  (DR7) galaxy sample represents the final set of galaxies observed
  using the original SDSS target selection criteria. We analyse the
  clustering of galaxies within this sample, including both the
  Luminous Red Galaxy (LRG) and Main samples, and also include the
  2-degree Field Galaxy Redshift Survey (2dFGRS) data. In total, this
  sample comprises $893\,319$ galaxies over 9\,100\,deg$^2$. Baryon
  Acoustic Oscillations are observed in power spectra measured for
  different slices in redshift; this allows us to constrain the
  distance--redshift relation at multiple epochs. We achieve a
  distance measure at redshift $z=0.275$, of $r_s(z_d)/D_V(0.275) =
  0.1390 \pm 0.0037$ (2.7\% accuracy), where $r_s(z_d)$ is the
  comoving sound horizon at the baryon drag epoch,
  $D_V(z)\equiv[(1+z)^2D_A^2cz/H(z)]^{1/3}$, $D_A(z)$ is the angular
  diameter distance and $H(z)$ is the Hubble parameter. We find an
  almost independent constraint on the ratio of distances
  $D_V(0.35)/D_V(0.2) = 1.736 \pm 0.065$, which is consistent at the
  1.1$\sigma$ level with the best fit $\Lambda$CDM model obtained when
  combining our $z=0.275$ distance constraint with the WMAP 5-year
  data. The offset is similar to that found in previous analyses of
  the SDSS DR5 sample, but the discrepancy is now of lower
  significance, a change caused by a revised error analysis and a
  change in the methodology adopted, as well as the addition of more
  data. Using WMAP5 constraints on $\Omega_b h^2$ and $\Omega_c h^2$,
  and combining our BAO distance measurements with those from the
  Union Supernova sample, places a tight constraint on $\Omega_m =
  0.286\pm0.018$ and $H_0 = 68.2\pm2.2\Hunit$ that is robust to
  allowing $\Omega_k \neq 0$ and $w \neq -1$. This result is
  independent of the behaviour of dark energy at redshifts greater
  than those probed by the BAO and supernova measurements.  Combining
  these data sets with the full WMAP5 likelihood constraints provides
  tight constraints on both $\Omega_k=-0.006\pm0.008$ and
  $w=-0.97\pm0.10$ for a constant dark energy equation of state.

\end{abstract}


\section{Introduction}  \label{sec:intro}

``What is the nature of dark energy?'' is one of the current key
questions in physical science. Distinguishing between competing
theories will only be achieved with precise measurements of the cosmic
expansion history and the growth of structure within it. Among current
measurement techniques for the cosmic expansion, Baryon Acoustic
Oscillations (BAO) appear to have the lowest level of systematic
uncertainty \citep{albrecht06}. 

BAO are a series of peaks and troughs, with a wavelength of
approximately $0.06\hompc$ that are present in the power spectrum of
matter fluctuations after the epoch of recombination, and on
large-scales. They occur because the primordial cosmological
perturbations excite sound waves in the relativistic plasma of the
early universe
\citep{silk68,peebles70,sunyaev70,bond84,bond87,holtzman89}. Radiation
pressure drives baryonic material away from the seed perturbations
until the ionised material recombines at redshift $z\simeq1000$. The
momentum of the baryonic material means that the motion continues for
a short time after recombination, until an epoch known as the
baryon-drag epoch. The wavelength of the BAO is related to the
comoving sound horizon at the baryon-drag epoch, which depends on the
physical densities of matter $\Omega_mh^2$ and of baryons
$\Omega_bh^2$ in the Universe. WMAP5 constraints on $\Omega_bh^2$ and
$\Omega_mh^2$ \citep{komatsu09} give that $r_s(z_d)\simeq153.5\,{\rm
  Mpc}$ (see Section~\ref{sec:interp} for details).

BAO occur on relatively large scales, which are still predominantly in
the linear regime at present day; it is therefore expected that BAO
should also be seen in the galaxy distribution
\citep{goldberg98,meiksin99,springel05,seo05,white05,eisenstein07}. We
can therefore use BAO as standard rulers to constrain the expansion of
the Universe if the comoving sound horizon at the baryon drag epoch is
known. The apparent size of the BAO measured from observations then
leads to measurements of the Hubble parameter and the angular diameter
distance \citep{seo03,blake03,hu03,matsubara04}.

The acoustic signature has now been convincingly detected at low
redshift \citep{percival01,cole05,eisenstein05,huetsi06} using the 2dF
Galaxy Redshift Survey (2dFGRS; \citealt{colless03}) and the Sloan
Digital Sky Survey (SDSS; \citealt{york00}). The detection has
subsequently been refined using more data and better techniques, and
is now producing competitive constraints on cosmological
models. \citet{tegmark06} analysed the Sloan Digital Sky Survey (SDSS)
Data Release 4 (DR4; \citealt{adelman06}) Luminous Red Galaxy (LRG)
sample. \citet{percival07a,percival07b} presented the power spectrum
of the Sloan Digital Sky Survey (SDSS) Data Release 5 (DR5;
\citealt{adelman07}) galaxy sample and considered the shape of the
power spectrum and measured the matter density using the BAO
features. \citet{percival07c} took this analysis a stage further by
fitting the SDSS data, combined with the 2dFGRS, with models of the
distance--redshift relation. \citet{gaztanaga08} and \citet{sanchez09}
have also analysed the SDSS DR6 \citep{adelman08} sample, obtaining
cosmological constraints from the radial and spherically averaged BAO
signal. In a recent analysis, \citet{kazin09} have calculated the
correlation function of the SDSS DR7 \citep{abazajian09} LRG sample,
and have shown that their results agree with those presented in our
paper. Two studies have also considered the clustering of the LRGs at
high redshift within the SDSS survey, using photometric redshifts to
estimate galaxy distances \citep{padmanabhan07,blake07}.

In this paper, we analyse the clustering of galaxies in the
spectroscopic SDSS DR7 sample, including both LRG and Main galaxy
samples, combined with the 2dFGRS, and measure the BAO signal in a
series of redshift slices. SDSS DR7 marks the final release of
galaxies observed using the standard SDSS targeting algorithm, and the
sample we analyse covers a solid angle of 7930\,deg$^2$, including a
7190\,deg$^2$ contiguous region in the North Galactic Cap. The Baryon
Oscillation Spectroscopic Survey (BOSS; \citealt{schlegel09a}), part
of the SDSS-III project, will adopt a different targeting algorithm,
focusing on galaxies and quasars at higher redshifts.

The observed amplitude of the large-scale galaxy clustering depends on
both galaxy colour and luminosity
\citep{tegmark04,zehavi05,swanson08}. Using the SDSS DR5 sample,
\citet{cresswell09} showed that for blue galaxies, the deviation in
the shape of the galaxy power spectrum from the linear matter power
spectrum at $k>0.1\hompc$ is a strong function of luminosity, while it
is almost constant for red galaxies. It is therefore difficult to
extract the underlying matter power spectrum from a galaxy power
spectrum measured for a population of galaxies where the distribution
of galaxy colours and luminosities changes with spatial location, such
as that provided by a magnitude-limited catalogue. In contrast, the
luminous red galaxy population, which comprises the high redshift part
of the sample analysed here, has a simpler relation with the matter
field, in that there is a single galaxy population to consider
\citep{reid08}. In a companion paper \citep{reid09}, we apply a
grouping algorithm to recover the halo power spectrum from the LRGs,
then calibrate the relation of the halo power spectrum to the linear
theory power spectrum using simulations.  We are then able to extract
cosmological information from the large scale shape of the power
spectrum in addition to the BAO signal, though the constraints are
more tightly embedded in the assumed cosmological framework.

BAO in the galaxy power spectrum are only weakly affected by the
effects of non-linear structure formation and scale-dependent galaxy
bias, because they are on such large scales. The primary consequence
is a damping on small scales, which can be well approximated by a
Gaussian smoothing \citep{bharadwaj96,crocce06,crocce08,
  eisenstein07,matsubara08a,matsubara08b}. The observed BAO, defined
as the ratio of the observed power spectrum $P_{\rm obs}$ to a smooth
fit to this power $P_{\rm nw}$, $BAO_{\rm obs}\equiv P_{\rm
  obs}/P_{\rm nw}$, are related to the original BAO in the linear
matter power spectrum $BAO_{\rm lin}$, defined similarly, by
\begin{equation}  \label{eq:bao_damp}
  BAO_{\rm obs}=G_{\rm damp}BAO_{\rm lin}+(1-G_{\rm damp}),
\end{equation} 
where $G_{\rm damp}=\exp(-\frac{1}{2}k^2D_{\rm damp}^2)$, and the
damping scale, $D_{\rm damp}$ is set to $10\mpcoh$ for redshift-space
power spectra at $z\simeq0.3$ \citep{eisenstein07}. This damping of
the linear power is a relatively benign effect as it does not affect
the positions of the BAO, although it does reduce the signal
available. Additional, more pernicious effects such as the mixing of
modes in the power spectrum, can generate shifts in the BAO position
\citep{crocce08}; for biased tracers, these offsets can be at the
percent level \citep{smith07}, and are therefore important as we wish
to make percent level distance measurements.

In our analysis, we measure BAO relative to a model that allows for
smooth changes in the underlying shape of the power spectrum, which
alleviates some of this shift. Physical models of BAO positions in
observed redshift-space power spectra relative to a such a fitted
smooth model \citep{crocce08,smith08,sanchez09,padmanabhan09b}, and
numerical simulations \citep{angulo07,seo08,kim09} suggest we should
expect residual shifts at the sub-percent level. These are below the
precision of current experiments: e.g. in this paper we present a BAO
distance scale measurement with 2.7\% accuracy. Therefore, we adopt a
procedure that allows for the damping as well as smooth changes in the
underlying shape of the power spectrum, but no more. The analysis of
future surveys, which will lead to tighter distance--redshift
constraints, will clearly also have to allow for non-linear effects,
either by physical modeling, simulations, or by using methods which
attempt to reconstruct the initial fluctuation field
\citep{eisenstein07,seo08,padmanabhan09a}.

The SDSS and 2dFGRS data are discussed in Sections~\ref{sec:data}
and~\ref{sec:slices}. The basic methodology, presented in
Section~\ref{sec:method}, is similar to that of \citet{percival07c},
although we have revised the calculation of the window function to
increase the computational speed. We also perform an extensive test of
the derived errors, running mock catalogues through our full analysis
pipeline to test the confidence intervals quoted
(Section~\ref{sec:bao_ln}). Results are presented in
Section~\ref{sec:results} and~\ref{sec:interp}, tested for robustness
in Section~\ref{sec:robust} and placed in a cosmological context in
Sections and~\ref{sec:cosmoparams}. A comparison with our DR5 analyses
is given in Section~\ref{sec:cmpr_dr5} and we finish with a discussion
in Section~\ref{sec:discuss}.

In this paper we use the standard cosmological parameters. For flat
$\Lambda$CDM models these are the Hubble constant $H_0$, the densities
of baryonic matter $\Omega_b$, cold dark matter $\Omega_c$, all matter
$\Omega_m$, and dark energy $\Omega_\Lambda$. Going beyond this simple
class of models, we use the equation of state of the dark energy $w$,
the curvature energy density $\Omega_k$ and total energy density
$\Omega_{\rm tot}$. When combining with information from the CMB, we
also consider some parameters that are not constrained by the BAO:
$\tau$ is the optical depth to re-ionization, $n_s$ is the scalar
spectral index, and $A_{05}$ is the amplitude of curvature
perturbations at $k=0.05\,{\rm Mpc}^{-1}$.

\section{The Data}  \label{sec:data}

The SDSS-I and SDSS-II projects used a 2.5m telescope \citep{gunn06},
to obtain imaging data in five passbands $u$, $g$, $r$, $i$ and $z$
\citep{fukugita96,gunn98}. The images were reduced
\citep{lupton01,stoughton02,pier03,ivezic04} and calibrated
\citep{lupton99,hogg01,smith02,tucker06}, and galaxies were selected
in two ways for follow-up spectroscopy. The main galaxy sample
\citep{strauss02} targeted galaxies brighter than $r=17.77$
(approximately $90$ per square degree, with a weighted median redshift
$z=0.10$). The DR7 sample \citep{abazajian09} used in our analysis
includes $669\,905$ main galaxies \citep{strauss02} with a median
redshift of $z=0.12$, selected to a limiting Galactic
extinction-corrected Petrosian magnitude $r<17.77$, or $r<17.5$ in a
small subset of the early data from the survey. The effect of the
inclusion of the early SDSS data is tested in
Section~\ref{sec:sample}. In addition, our sample includes $80\,046$
Luminous Red Galaxies (LRGs; \citealt{eisenstein01}), which form an
extension of the SDSS spectroscopic survey to higher redshifts
$0.2<z<0.5$. Of the main galaxies, $30\,530$ are also classified as
LRGs and are intrinsically luminous with ${\rm M}_{^{0.1}r}<-21.8$,
where ${\rm M}_{^{0.1}r}$ is the Galactic extinction and K-corrected
$r$-band absolute galaxy magnitude. We apply this requirement to all
of our LRGs, so our sample includes $110\,576$ LRGs in total, with a
weighted median redshift of $z=0.31$. Although the main galaxy sample
contains significantly more galaxies than the LRG sample, the LRG
sample covers more volume. Redshift distributions for these samples
are shown in figure~2 of \citet{percival07b}.  In our default analysis
we use SDSS Petrosian magnitudes calibrated using the
``uber-calibration'' method \citep{padmanabhan08}, although we test
against data calculated using the original calibration methodology
\citep{tucker06}. Where specified, we have K-corrected the galaxy
luminosities using the methodology outlined by
\citet{blanton03a,blanton03b}. Further details of the cuts applied to
the data can be found in \citet{percival07b}.

Due to the finite size of the fibers, spectra cannot be obtained for
both objects in a pair closer than 55\,arcsec, within a single
spectroscopic tile. Tiling \citep{blanton03a} deals with this to some
extent by allowing plate overlaps to provide multiple observations of
crowded regions. Even so, not all galaxies in such regions which meet
the target selection criteria could be observed. \citet{zehavi02}
corrected for this undersampling by assigning the redshift of the
nearest observed galaxy to a galaxy which was not observed due to
crowding, and showed that this provides sufficient correction for
large-scale structure studies.  We apply this correction in the
present work, and test it to show that our results are insensitive to
this in Section~\ref{sec:zspace}.

In order to increase the volume covered at redshift $z<0.3$, we
include $143\,368$ galaxies from the 2dFGRS sample. These galaxies,
selected to an extinction-corrected magnitude limit of approximately
$b_J=19.45$ \citep{colless03} from regions of sky not covered by the
SDSS sample, cover two contiguous regions totalling
$\sim$1200deg$^2$. They do not include the 2dFGRS random fields, a set
of 99 random 2~degree fields spread over the full southern Galactic
cap, as these would complicate the window function. The galaxies cover
$0<z<0.3$, with a weighted median at $z=0.17$. The redshift
distribution of the sample was analysed as in \citet{cole05} for
$0<z<0.3$, and we use the same synthetic catalogues to model the
unclustered expected galaxy distribution within the reduced sample.

We assume that each galaxy is biased with a linear deterministic bias
model, and that this bias depends on ${\rm M}_{^{0.1}r}$ according to
\citet{tegmark04} and \citet{zehavi05}. All galaxies were weighted
using this model so the fluctuation amplitudes match those of $L_*$
galaxies, where $L_*$ was calculated separately for the SDSS and
2dFGRS. We include an extra normalisation factor to the 2dFGRS galaxy
bias model to correct the relative bias of $L_*$ galaxies in the
different surveys. This was calculated by matching the normalisation
of the 2dFGRS and SDSS bias-corrected power spectra for
$k<0.1\hompc$. In principle, we could have added information on galaxy
bias from the BAO, since the small-scale damping (see
Eq.~\ref{eq:bao_damp}) depends on how strongly nonlinear the
underlying dark matter density fluctuations are. As we show in
Section~\ref{sec:Ddamp}, this information is limited for the current
data, but future surveys may be able to exploit changes in this
damping as a function of galaxy properties, such as colour and
luminosity.

\section{Splitting into sub-samples}  \label{sec:slices}

\begin{table}
\begin{center}
\begin{tabular}{cccrcr}
\hline
SLICE & $z_{\rm min}$ & $z_{\rm max}$ & $N_{\rm gal}$ & $V_{\rm eff}$ & $\bar{n}$ \\
\hline
1 & 0.0 & 0.5 & 895\,834 & 0.42 & 128.1 \\ 
2 & 0.0 & 0.4 & 874\,330 & 0.38 & 131.2 \\  
3 & 0.0 & 0.3 & 827\,760 & 0.27 & 138.3 \\
4 & 0.1 & 0.5 & 505\,355 & 0.40 & 34.5  \\
5 & 0.1 & 0.4 & 483\,851 & 0.36 & 35.9  \\
6 & 0.2 & 0.5 & 129\,045 & 0.27 & 1.92  \\ 
7 & 0.3 & 0.5 &  68\,074 & 0.15 & 0.67  \\ 
\hline
\end{tabular}
\end{center}
\caption{Parameters of the redshift intervals analysed. $V_{\rm eff}$ is 
  given in units of $h^{-3}{\rm Gpc}^3$, and was calculated as in 
  Eq.~(\ref{eq:veff}) using an effective power spectrum amplitude of
  $\bar{P}=10^4h^{-3}{\rm Mpc}$, appropriate on scales $k\sim0.15\hompc$ 
  for a population with bias $b=1.7$. The average galaxy number density 
  in each bin $\bar{n}$ is in units of $10^{-4}(\mpcoh)^3$.}
\label{tab:slices} 
\end{table}

In order to probe the distance--redshift relation in detail, ideally
we would analyze BAO measured in many independent redshift
slices. However, if the slices are too narrow in redshift, then there
is insufficient signal and the BAO cannot be recovered with sufficient
accuracy to give a likelihood with close to a Gaussian distribution
(see the discussion in Section~\ref{sec:bao_ln}). If the slices are
too wide, or too many overlapping slices are chosen, the covariance
matrix becomes close to singular, potentially leading to numerical
instability. In order to balance these competing requirements, we have
chosen to analyse the redshift slices presented in
Table~\ref{tab:slices}. The power spectra will be correlated, and
these correlations, together with correlations of $P(k)$ values at
different $k$ within each redshift slice, are included in the
covariance matrices in our analysis. Note that we include slice 7, for
which the effective volume is relatively small, because of the
interesting redshift range covered.

As well as giving the redshift limits of the slices in
Table~\ref{tab:slices}, we also give the number of galaxies in each
including both the 2dFGRS and the SDSS, and the effective volume,
calculated from the integral \citep{FKP}
\begin{equation}  \label{eq:veff}
  V_{\rm eff} = \int d^3r\left[\frac{\bar{n}({\bf r})\bar{P}}
    {1+\bar{n}({\bf r})\bar{P}}\right]^2,
\end{equation}
where $\bar{n}({\bf r})$ is the observed comoving number density of
the sample at location ${\bf r}$ and $\bar{P}$ is the expected power
spectrum amplitude. To calculate $V_{\rm eff}$ for our redshift
slices, distances were calculated assuming a fiducial flat
$\Lambda$CDM cosmology with $\Omega_m=0.25$. For the numbers given in
Table~\ref{tab:slices}, we fix $\bar{P}=10^4h^{-3}{\rm Mpc}^3$,
appropriate on scales $k\sim0.15\hompc$ for a population with bias
$b=1.7$. For comparison, \citet{eisenstein05} analyse a sample with
$V_{\rm eff} = 0.13\,h^{-3}{\rm Gpc}^3$, approximately a third of the
effective volume of slice 1.

We fit models to three sets of power spectra:
\begin{enumerate}
\item We fit a single power spectrum for the SDSS LRG sample covering
  $0.15<z<0.5$.
  \label{it1}
\item We fit three power spectra for slices 1, 3 and 6 approximately
  corresponding to the procedure adopted by
  \citet{percival07c}. Although we now use slices constrained by
  redshift rather than galaxy type, the $0<z<0.3$ slice is dominated
  by SDSS main galaxies, while the $0.2<z<0.5$ slice is dominated by
  LRGs.
  \label{it2}
\item We fit six power spectra for slices 2$\to$7, which allows a test
  of the distance--redshift relation at greater resolution.
  \label{it3}
\end{enumerate}
We consider option \ref{it1} to tie in with the analysis presented by
\citet{reid09}, and to demonstrate the effect of collapsing the
clusters in redshift-space where we try to reconstruct the halo power
spectrum. Option \ref{it2} is close to the approach of
\citet{percival07c}, where the SDSS main galaxies and 2dFGRS galaxies
were analysed separately from the SDSS LRGs. Option \ref{it3} allows
us to see if there is more information available beyond measuring the
distance--redshift relation at two redshifts. The slices do overlap in
redshift, but we will properly take into account the covariance
between the results when we fit to cosmological parameters.

\section{Basic methodology}  \label{sec:method}

Power spectra were calculated for each catalogue using the Fourier
method of \citet{FKP}, as applied by \citet{percival07b}. In this
method a weighted galaxy over-density field is defined and Fourier
transformed, then the spherically averaged power is measured. We use
the luminosity dependent galaxy weights advocated by \citet{PVP}, as
described in Section~\ref{sec:data}. To construct the over-density
field, we need to quantify the expected spatial distribution of
galaxies, in the absence of clustering. The standard method for this
is to use an unclustered random catalogue, which matches the galaxy
selection. To calculate this random catalogue, we fitted the redshift
distributions of the galaxy samples with a spline fit \citep{press92},
and the angular mask was determined using a routine based on a {\sc
  HEALPIX} \citep{gorski05} equal-area pixelization of the sphere
\citep{percival07b}. \citet{percival07b} used a random catalogue
containing ten times as many points as galaxies. For the sparse LRGs,
this approach induces significant shot noise, so we now use one
hundred times as many random points as LRGs. We have also increased
the resolution at which the radial distribution of galaxies is
quantified, now using a spline fit \citep{press92} with nodes
separated by $\Delta z=0.0025$. As an alternative to this radial
selection, we could have simply adopted the redshift of a randomly
chosen galaxy for each of our points in the random catalogue. In
Section~\ref{sec:zfit} we show that these two possibilities give
consistent results.

Galaxy redshifts were converted to distances using a fiducial
cosmology (flat $\Lambda$CDM model with $\Omega_m=0.25$). For each
distance--redshift model to be tested, we do not recalculate the power
spectrum, but instead change the interpretation of the power spectrum
computed assuming the fiducial $\Lambda$CDM galaxy distances. We do
this through a window function, which relates the true and measured
power spectra. This follows the procedure adopted by
\citet{percival07c}, but we now use a revised, computationally less
intensive method for calculating the windows, as described in
Appendix~\ref{app:win}.

A model of the BAO was created by fitting a linear matter power
spectrum, calculated using {\sc CAMB} \citep{lewis00}, which
numerically solves the Boltzman equation describing the physical
processes in the Universe before the baryon-drag epoch, with a cubic
spline to remove the broad shape of the power, leaving the
oscillations. The theoretical BAO were then damped with a Gaussian
model as in Eq.~(\ref{eq:bao_damp}), following the simulation results
of \citet{eisenstein07}. For our default fits, we assume that the
damping scale $D_{\rm damp}=10\mpcoh$ \citep{eisenstein07}, but we
also consider fits where this scale is varied
(Section~\ref{sec:Ddamp}). As discussed in Section~\ref{sec:intro}, we
do not attempt to correct for any shift induced by non-linear physics,
because they are expected to be at a level below our statistical
error.

The power spectrum measured from the data was fitted by a model
constructed by multiplying this BAO model with a cubic spline
\citep{press92}, which enables the power spectrum model to match the
overall shape of the data power spectrum. Each power spectrum model
was then convolved with a window function that corrects for both the
survey geometry and the differences between our fiducial cosmological
model used to convert redshift to distances and the cosmological model
to be tested (see Appendix~\ref{app:win}). The free parameters of the
model are the nine nodes of the cubic spline fixed empirically at
$k=0.001$, and $0.025\le k\le0.375$ with $\Delta k=0.05$, and the
parametrisation of $D_V(z)$ used to calculate the correct window
function. The spline nodes were refitted for every cosmology (or
$D_V(z)$) tested. A power spectrum model with this spline node
separation was tested by fitting many mock power spectra by
\citet{percival07a} and was shown to match these without leaving
significant residuals in the measured ``shift'' between BAO in the
model and data power spectra. This approach was also considered by
\citet{sanchez08}, who found that it did not induce a bias in the
recovered BAO constraints.

For a redshift survey in a thin shell, the position of the BAO
approximately constrains $d_z\equiv r_s(z_d)/D_V(z)$, where $r_s(z_d)$
is the comoving sound horizon at the baryon drag epoch,
$D_V(z)\equiv[(1+z)^2D_A^2cz/H(z)]^{1/3}$
\citep{eisenstein05,percival07c}, $D_A$ is the angular diameter
distance, and $H(z)$ is the Hubble parameter. We see that, although
our power spectrum fitting procedure measures $D_V(z)$ for a fixed BAO
model, we should consider the constraints as measurements of $d_z$,
with $r_s(z_d)$ calculated for the flat $\Lambda$CDM model for which
we created the BAO model, $r_s(z_d)=111.4\mpcoh=154.7\,{\rm Mpc}$,
using equation~6 of \citet{eisenstein98}, and assuming $h=0.72$,
$\Omega_bh^2=0.0223$, and $\Omega_m=0.25$. This value of $r_s(z_d)$ is
only used to index this model: as described above, the actual BAO
model was calculated from a power spectrum predicted by {\sc CAMB}. If
the constraints provided in this paper are to be used to constrain a
set of models where $r_s(z_d)$ for this fiducial model is calculated
in a different way (i.e. not using equation~6 of
\citealt{eisenstein98}), then our constraints should be adjusted to
match.

The comoving distance--redshift relation is modelled as a cubic spline
in the parameter $D_V(z)$. We consider models for $D_V(z)$ with two
nodes at $z=0.2$ and $z=0.35$, or with four nodes at $z=0.1$, $z=0.2$,
$z=0.3$, $z=0.45$. Results are presented as constraints on $d_z$. The
error between cubic spline fits to $D_V(z)$ with two nodes at $z=0.2$
and $z=0.35$, to the $\Lambda$CDM distance--redshift relations was
shown in figure~1 of \citet{percival07c}, and is $<1\%$ for a flat
$\Lambda$CDM cosmology with $\Omega_m=0.25$ at $z\ge0.15$.

\begin{figure}
  \centering
  \resizebox{0.9\columnwidth}{!}{\includegraphics{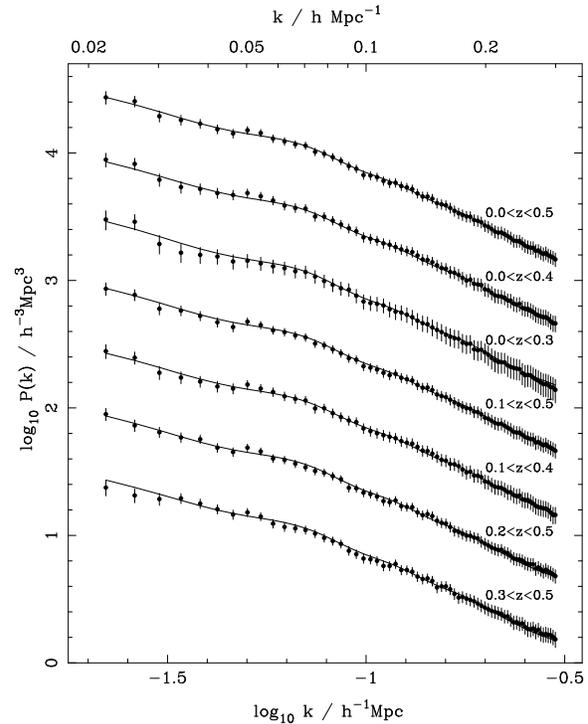}}
  \caption{Average power spectra recovered from the Log-Normal
    catalogues (solid lines) compared with the data power spectra
    (solid circles with 1-$\sigma$ errors) for the six samples in
    Table~\ref{tab:slices}. The errors on the data were calculated
    from the diagonal elements of the covariance matrix calculated
    from these log-normal catalogues. The power spectra have been
    offset by 0.5dex for clarity, with the upper power spectrum having
    the correct normalisation.}
  \label{fig:pk_vs_ln}
\end{figure}
Power spectra are presented for the redshift slices described in
Section~\ref{sec:slices} in Fig.~\ref{fig:pk_vs_ln}, for $70$ band
powers equally spaced in $0.02<k<0.3\hompc$. We see that the power
spectra from the different redshift intervals are remarkably
consistent, with $P(k)$ decreasing almost monotonically to small
scales. 

In order to calculate the covariances between the data, we have
created 10\,000 Log-Normal (LN) density fields \citep{coles91,cole05}
from which we have drawn overlapping catalogues for each of our 7
redshift slices. Catalogues were calculated on a $(512)^3$ grid with
box length $4000\mpcoh$. Unlike N-body simulations, these mock
catalogues do not model the growth of structure, but instead return a
density field with a log-normal distribution, similar to that seen in
the real data. The window functions for these catalogues were matched
to that of the 2dFGRS+SDSS catalogue with the original
calibration. The input power spectrum was a cubic spline fit matched
to the data power spectra (i.e. the smooth part of our standard
model), multiplied by our default damped $\Lambda$CDM BAO model
calculated using {\sc CAMB} \citep{lewis00}. The LN power spectra were
used to determine a covariance matrix between slices and for different
band powers in each slice, assuming that the band-powers were drawn
from a multi-variate Gaussian distribution. Average recovered power
spectra for each redshift interval are compared with the data power
spectra in Fig.~\ref{fig:pk_vs_ln}. Clearly the general shape of the
average power spectra of the LN catalogues is well matched to that
recovered from the data. Using the inverse of this covariance matrix,
we estimate the likelihood of each model assuming that the power
spectra band-powers for $0.02<k<0.3\hompc$ were drawn from a
multi-variate Gaussian distribution.

\section{Testing the analysis method with mock data}  \label{sec:bao_ln}

\subsection{The model fit}

We now consider using a subset of our LN catalogues to test our
analysis procedure. For 1000 of the mock catalogues, we fit
spline$\times$BAO models to extract distance constraints from the BAO,
as described in Section~\ref{sec:method}. A small average shift of
1.3\% in the BAO scale was recovered between the power recovered from
the LN catalogues, and the input power spectrum used to create
them. If we correct the 1000 power spectra measured from the LN
catalogues by multiplying each power spectrum by the expected power
divided by the average recovered power spectrum, the average shift
drops below $0.3\%$, well within 1$\sigma$.

To test the origin of the observed 1.3\% shift, we have also drawn
1000 power spectrum realisations from a multi-variate Gaussian
distribution with covariance and mean matched to those of the
data. These mock catalogues were fitted using the procedure described
in Section~\ref{sec:method}. No shift in the BAO position was found
from the fits to these catalogues, within the statistical limits of
the analysis ($\sim0.3\%$). The distribution of recovered distance
constraints was well matched to that recovered from fitting the
corrected LN power spectra. Thus the 1.3\% shift described above must
be due to the LN procedure itself. The expected shift is dependent on
the statistic used to measure the BAO position. The Log-Normal
correlation function $\xi_{LN}$, and Gaussian correlation function
$\xi_G(r)$ of a field with the same power spectrum but with Gaussian
statistics, are related by $1+\xi_{LN}(r) = \exp[\xi_{G}(r)]$. If we
had used the peak in the correlation function as our standard ruler
then, for the LN catalogues, we would have expected no BAO
shift. However, the same is not true of our BAO $\times$ spline model
fitting procedure, which fits the BAO in the power spectrum over a
range of scales.

Numerical simulations offer a better way to model the true Universe,
and recent results from simulations show that we should expect a less
significant shift between the BAO positions in the linear matter and
galaxy power spectra than the 1.3\% shift found for the LN catalogues
\citep{seo03,springel05,seo07,angulo07}. The exact shift required for
the catalogues we analyse is not well constrained by these simulation
results, and we consequently do not alter our analysis to include such
a shift.

\subsection{The likelihood surface}

\begin{table}
  \begin{tabular}{ccccc}
    \hline
    $-2\ln{\cal L}/{\cal L}_{\rm true}$ & 
    \multicolumn{4}{c}{fraction of samples}\\
    & \multicolumn{2}{c}{3 slices} & \multicolumn{2}{c}{6 slices}\\
    & standard & revised & standard & revised\\
    \hline
    $<$2.3 & 0.579 & 0.666 & 0.551 & 0.667\\ 
    $<$6.0 & 0.892 & 0.946 & 0.862 & 0.948\\ 
    $<$9.3 & 0.966 & 0.983 & 0.955 & 0.981\\ 
    \hline
  \end{tabular}
  \caption{Fraction of fits to the Log-Normal power spectra in which the 
    ratio of the likelihood maximum and the likelihood for the true 
    cosmological model is less than the given limit. For a Gaussian 
    likelihood, these limits correspond to 68\%, 95\% and 99\% confidence 
    intervals. We show results 
    where we have corrected the errors as described in the text by 
    multiplying the band-power errors by $1.14$ for three redshift slices, 
    and $1.21$ for six redshift slices.}
  \label{tab:LN_dist}
\end{table}
We use the Gaussian and LN power spectra samples to assess the nature
of the likelihood for the BAO scale recovery. We consider fits to
either three or six power spectra as described in
Section~\ref{sec:slices}, parametrising $D_V(z)$ with a cubic spline
with two non-zero nodes at $z=0.2$ and $z=0.35$. For each of the 1000
fits, we have measured the difference between the maximum likelihood
value and the likelihood at the parameters of the true cosmological
model. The fraction of samples with $-2\ln{\cal L}/{\cal L}_{\rm
  true}<2.3,\,6.0,\,9.3$, corresponding to 68\%, 95\% and 99\%
confidence intervals, are given in Table~\ref{tab:LN_dist}. We find
that in order to match the expected numbers of samples with
likelihoods within the standard 1$\sigma$ Gaussian confidence
intervals, we must increase the errors on the power spectrum band
powers by $14\pm2$\% if we fit to three power spectra. For fits to six
power spectra, we must increase the errors by $21\pm2$\% in order to
match the expected 1$\sigma$ Gaussian confidence intervals. Although
in this paper we do not consider fitting to a single power spectrum,
we have repeated this analysis for BAO fits to the LRG sample of
\citet{reid09}, and find that we must increase the errors on the power
spectrum band powers by $10\pm2$\% to match the expected confidence
intervals.

Because the same increase in the confidence intervals is required for
both LN and Gaussian mock catalogues, this change must be caused by
the methodology of fitting BAO, rather than the Gaussian to Log-Normal
density field transition. In fact, we believe that it is caused by the
non-Gaussian nature of the likelihood surface. We should expect the
likelihood surface to be non-Gaussian to some extent in any case
because there is a minimum in the likelihood where the observed and
model BAO are perfectly out of phase in $k$-space: this represents the
worst possible match between data and model. Adjusting the covariance
matrix to match the distribution of best-fit distance-scales to the
expected 68\% confidence interval does not quite match the 95\% or
99\% confidence intervals, although it corrects for most of the
difference. This shows that the confidence intervals cannot perfectly
match those for a Gaussian distribution.

To test this further, we have created a set of 1000 Gaussian power
spectrum realisations with errors that are 10\% of those in our
standard sample. For these catalogues, the distribution of best-fit
$D_V(z)$ matches that expected from the likelihood distribution under
Gaussian assumptions. No correction is required, and the likelihood
distribution is much closer to that for a multi-variate Gaussian
distribution around the likelihood maximum. Thus the requirement to
increase the errors on the data disappears when we fit less noisy
data, as we would expect if it is caused by fitting noisy data, which
is giving a non-Gaussian likelihood surface.

\begin{figure}
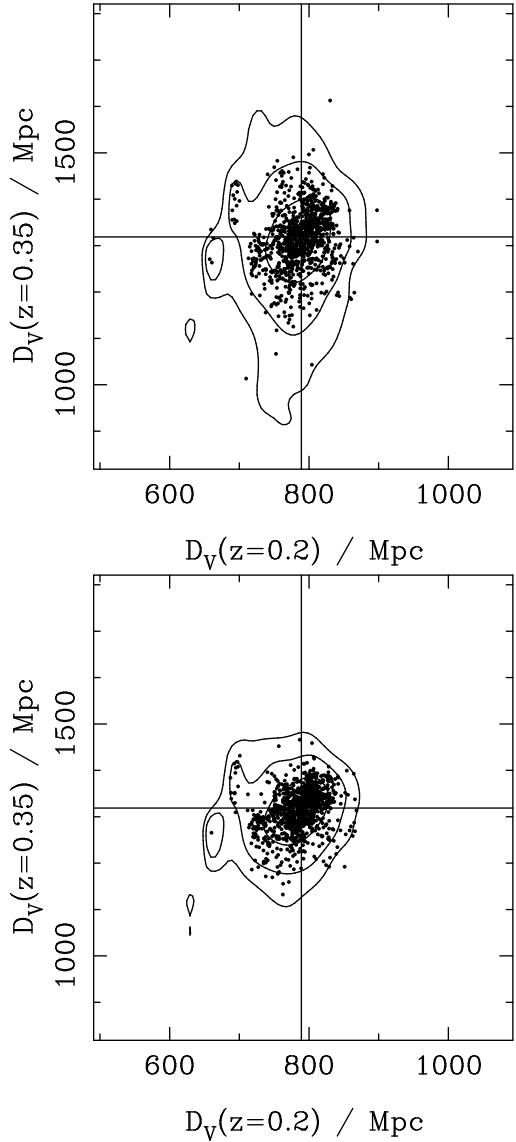

  \centering
  \resizebox{0.8\columnwidth}{!}{\includegraphics{like_LN_2par_3bao.ps}}
  \resizebox{0.8\columnwidth}{!}{\includegraphics{like_LN_2par_6bao.ps}}
  \caption{Average likelihood contours recovered from the analysis of
    three power spectra (top panel) and six power spectra (bottom
    panel) measured from 1000 Log-Normal density fields. Contours are
    plotted for $-2\ln{\cal L}=2.3,\,6.0,\,9.2$, corresponding to
    two-parameter confidence of 68\%, 95\% and 99\% for a Gaussian
    distribution. Contours were calculated after increasing the errors
    on the power spectrum band-powers as described in the text. Solid
    circles mark the locations of the likelihood maxima closest to the
    true cosmology. We have plotted the likelihood surface as a
    function of $D_V(z) / {\rm Mpc}$, for fixed $r_s(z_d)=154.7\,{\rm
      Mpc}$, to show distance errors if the comoving sound horizon is
    known perfectly. The values of $D_V$ for our input cosmology are
    shown by the vertical and horizontal solid lines.}
 \label{fig:like_LN}
\end{figure}
The average likelihood surfaces measured from our 1000 fits to sets of
three power spectra and six power spectra drawn from LN catalogues are
shown in Fig.~\ref{fig:like_LN}. We also plot the centre of the local
likelihood maxima nearest to the input cosmological parameters for
each model. The fractions of points within each contour are given in
Table~\ref{tab:LN_dist}: the errors on the power spectrum band powers
have been adjusted for each plot as described above so that $\sim$68\%
of the points lie within the $-2\ln{\cal L}=2.3$ contour.

\section{Results}  \label{sec:results}

\begin{figure}
  \centering
  \resizebox{0.9\columnwidth}{!}{\includegraphics{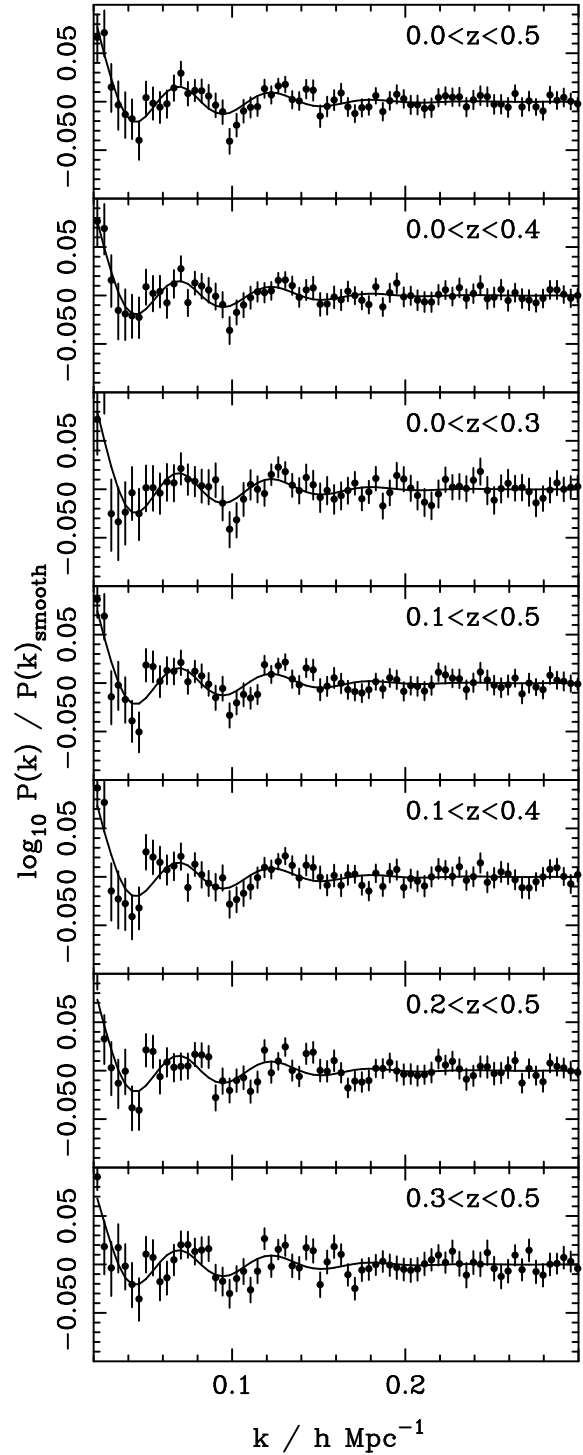}}
  \caption{BAO recovered from the data for each of the redshifts
    slices (solid circles with 1-$\sigma$ errors). These are compared
    with BAO in our default $\Lambda$CDM model (solid
    lines).}
  \label{fig:bao_data}
\end{figure}

\begin{figure}
\centering
\resizebox{0.9\columnwidth}{!}{\includegraphics{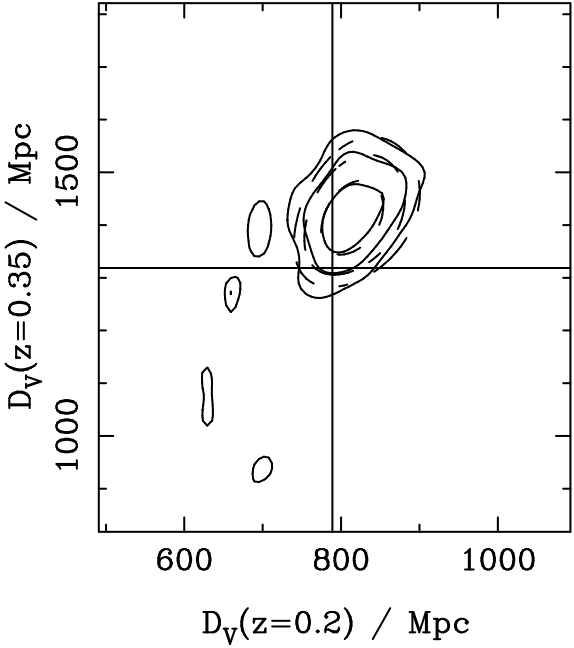}}
\caption{Likelihood contour plots for fits of two $D_V(z)$ cubic
  spline nodes at $z=0.2$ and $z=0.35$, calculated for our default
  analysis using six power spectra, uber-calibration, a fixed BAO
  damping scale of $D_{\rm damp}=10\mpcoh$, and for all SDSS and
  non-overlapping 2dFGRS data. Solid contours are plotted for
  $-2\ln{\cal L}/{\cal L}_{\rm true}<2.3,\,6.0,\,9.3$, which for a
  multi-variate Gaussian distribution with two degrees of freedom
  correspond to 68\%, 95\% and 99\% confidence intervals. Likelihoods
  were adjusted to match these Gaussian confidence intervals as
  described in Section~\ref{sec:bao_ln}. We have plotted the
  likelihood surface as a function of $D_V(z) / {\rm Mpc}$, for fixed
  $r_s(z_d)=154.7\,{\rm Mpc}$, to show distance errors if the comoving
  sound horizon is known perfectly. We also show a multi-variate
  Gaussian fit to this likelihood surface (dashed contours). The
  values of $D_V$ for a flat $\Lambda$CDM cosmology with
  $\Omega_m=0.25$, $h=0.72$, \& $\Omega_bh^2=0.0223$ are shown by the
  vertical and horizontal solid lines.}
\label{fig:like_data3}
\end{figure}

Baryon Acoustic Oscillations are observed in the power spectra
recovered from all redshift slices of the SDSS+2dFGRS sample described
in Section~\ref{sec:slices}, and are shown in Fig.~\ref{fig:bao_data},
where we plot the measured power spectra divided by the spline
component of the best-fit model. In our default analysis we fit power
spectra from six redshift slices as described in
Section~\ref{sec:slices}, using a spline for $D_V(z)$ with two nodes
at $z=0.2$ and $z=0.35$. We assume a fixed BAO damping scale of
$D_{\rm damp}=10\mpcoh$ and fit to all SDSS and non-overlapping 2dFGRS
data. The effect of these assumptions is considered in
Section~\ref{sec:robust}. The resulting likelihood surface is shown in
Fig.~\ref{fig:like_data3} as a function of $D_V(z) / {\rm Mpc}$, for
fixed $r_s(z_d)=154.7\,{\rm Mpc}$, to show distance errors if the
comoving sound horizon is known perfectly. The constraints should be
considered measurements of $r_s(z_d)/D_V(z)$ (see
Section~\ref{sec:method}). Fig.~\ref{fig:like_data3} reveals a
dominant likelihood maximum close to the parameters of a $\Lambda$CDM
cosmology with $\Omega_m=0.25$, $h=0.72$, \&
$\Omega_bh^2=0.0223$. There are also weaker secondary maxima at lower
$D_V(0.2)$, which are considered further in
Section~\ref{sec:secondaries}.  The significance of detection of BAO
corresponds to $\Delta \chi^2 = 13.1$, which is approximately
$3.6\sigma$. As this is relative to an arbitrary smooth model, this
test is more general, and hence the significance cannot be directly
compared with results presented by \citet{eisenstein05}.

We have matched the likelihood surface shown in
Fig.~\ref{fig:like_data3} around the dominant maximum to a
multi-variate Gaussian model. Using this Gaussian fit, we find that
the best fit model has
\begin{eqnarray}  
  d_{0.2}  &=& 0.1905\pm0.0061 \;\;(3.2\%), \nonumber \\
  d_{0.35} &=& 0.1097\pm0.0036 \;\;(3.3\%), \label{eq:DV_constraints}
\end{eqnarray}
where $d_z\equiv r_s(z_d)/D_V(z)$. These results are correlated with
correlation coefficient $r=0.337$. For a cosmological
distance--redshift model with $\hat{d}_z$ the likelihood can be well
approximated by a multi-variate Gaussian with covariance matrix
\begin{equation}
  C \equiv \left(\begin{array}{cc} 
    \left\langle \Delta d_{0.2}\Delta d_{0.2}\right\rangle & 
    \left\langle \Delta d_{0.2}\Delta d_{0.35}\right\rangle \\
    \left\langle \Delta d_{0.35}\Delta d_{0.2}\right\rangle & 
    \left\langle \Delta d_{0.35}\Delta d_{0.35}\right\rangle, \\
   \end{array}\right),
\end{equation}
where $\Delta d_z\equiv d_z-\hat{d}_z$. $C$ has inverse
\begin{equation}  \label{eq:DV_invcov}
  C^{-1} = \left(\begin{array}{cc} 
    30124 & -17227 \\
   -17227 &  86977 \\
   \end{array}\right).
\end{equation}

Without correcting the covariance matrix using the results from
fitting to the LN power spectra as described in
Section~\ref{sec:bao_ln}, the original average errors on $d_{0.2}$,
and $d_{0.35}$ were $0.0051$ and $0.0029$, 16\% and 24\% lower than
those in Eq.~(\ref{eq:DV_constraints}), respectively. Compare with the
band power errors which were increased by 21\%, and we see that there
is not a direct relation between changes in the band power errors and
errors on $d_z$, because of the non-linear nature of the fit.

We diagonalise the covariance matrix of $d_{0.2}$ and $d_{0.35}$ to
get quantities $x$ and $y$
\begin{equation}  \label{eq:xy}
  \left(\begin{array}{c}
      x \\
      y
    \end{array}\right)
  \equiv
  \left(\begin{array}{cc}
      1  & 1.76 \\
      -1 & 1.67
    \end{array}\right)
  \left(\begin{array}{c}
      d_{0.2} \\
      d_{0.35}
    \end{array}\right),
\end{equation}
which gives
\begin{eqnarray}
  x &=& 0.3836  \pm 0.0102\\
  y &=& -0.0073 \pm 0.0070.
\end{eqnarray}
The distance ratio $f\equiv D_V(0.35)/D_V(0.2)$ is given by
\begin{equation}
  f=\frac{1.67-1.76y/x}{1+y/x}\simeq1.67-8.94y,
\end{equation}
where the last approximation neglects the small variations around the
best-fit value of $x=0.3836$; these would come to $0.002$ in $f$,
which is well within the errors. Thus, $x$ is a measurement of
distance for the concordance cosmology and $y$ is the deviation from
the concordance distance ratio: $x$ is measured to about 2.7\%.  $y$
is consistent with zero to within about $1\sigma$.

To high accuracy, the constraint $x$ can be written as a constraint on
the distance to some redshift $0.2<z<0.35$. In fact,
$r_s(z_d)/D_V(0.275)$ predicts $x = d_{0.2}+1.76d_{0.35}$ to a
peak-to-peak precision of 0.04\% over the range $0.05<\Omega_m<1$
(assuming a flat cosmology with $w=-1$). Thus, we can quote the $x$
measurement as a measurement of $d_{0.275}$ and quote the $y$
measurement as a statistically independent measure of $f$.

For the best-fit solution we have $d_{0.275} = 0.362x$, giving
\begin{equation} \label{eq:d0.275}
  d_{0.275} = 0.1390 \pm 0.0037 (2.7\%).
\end{equation}
We also have the statistically independent constraint
\begin{equation} \label{eq:frat}
  f \equiv D_V(0.35)/D_V(0.2)=1.736\pm0.065.
\end{equation}
$f=1.67$ for our $\Lambda$CDM concordance cosmology, while SCDM with
$\Omega_m=1$, $\Omega_\Lambda=0$ has $f=1.55$, which is only
2.9$\sigma$ from this result. Our constraint from the distance ratio
only separates the concordance model from $\Omega_m=1$ at 1.8$\sigma$,
i.e., it is not a strong cosmological constraint, compared with the
constraint on $d_{0.275}$.

\section{Cosmological Interpretation}  \label{sec:interp}

We now consider how our constraints can be mapped into the standard
basis of cosmological parameters. From equation~6 of
\citet{eisenstein98}, the sound horizon can be approximated, around
the WMAP5 best-fit location \citep{komatsu09} as
\begin{equation}  \label{eq:sound-horizon}
  r_s(z_d) = 153.5
  \left(\Omega_b h^2\over 0.02273\right)^{-0.134}
  \left(\Omega_m h^2\over 0.1326\right)^{-0.255} 
  {\rm\ Mpc}.
\end{equation}
Setting $r_{s,fid}=153.5\,{\rm Mpc}$, and using Eq.~(\ref{eq:d0.275})
we have
\begin{displaymath}
  D_V(0.275) = (1104 \pm 30) [r_s(z_d)/r_{s,fid}(z_d)] {\rm\ Mpc}
\end{displaymath}
\begin{equation}  \label{eq:DV0275}
    = (1104 \pm 30) 
    \left(\Omega_b h^2\over 0.02273\right)^{-0.134}
    \left(\Omega_m h^2\over 0.1326\right)^{-0.255} {\rm\ Mpc},
\end{equation}
and $f = 1.736\pm0.065$ as our two statistically independent
constraints.

The constraint on $D_V(0.275)$, combined with a measurement of
$\Omega_m h^2$ from WMAP5 \citep{hinshaw09,dunkley09,komatsu09}, is
enough to measure $\Omega_m$ and $H_0$ given information about the
distance scale from $z=0$ to $z=0.275$. If the distance measure were
at $z=0$, then we would have a standard ruler defined by the CMB with
which we could measure $H_0$, and combining this with $\Omega_m h^2$
would yield $\Omega_m$. In practice, one has to include the small
corrections to $D_V(0.275)$ that arise from the low-redshift
cosmology. Noting that $D_V(0.275)=757.4\mpcoh$ for a flat
$\Omega_m=0.282$ $\Lambda$CDM cosmology, we can write
$h=\sqrt{\Omega_mh^2}/\sqrt{\Omega_m}$, and solve
\begin{eqnarray} 
  \Omega_m &=& \left(0.282\pm0.015\right)
    \left(\frac{\Omega_mh^2}{0.1326}\right)^{0.49}\nonumber\\
    & & \hspace{1cm}
    \times
    \left(\frac{D_V(z=0.275,\Omega_m=0.282)}{D_V(z=0.275)}\right)^2,
    \label{eq:om_m1}
\end{eqnarray}
where we have dropped the dependence of the sound horizon on
$\Omega_bh^2$, which the WMAP5 data already constrains to 0.5\%, 5
times below our statistical error.

We can perturb the ratio of distances around the best-fit
$\Omega_m=0.282$, to give
\begin{displaymath}
  \frac{D_V(z=0.275)}{D_V(z=0.275,\Omega_m=0.282)}
\end{displaymath}
\begin{equation}
\hspace{0.5cm}
  = \left(\frac{\Omega_m}{0.282}\right)^{-0.077}
    \left[1-0.108\Omega_k-0.099(1+w)\right].
\end{equation}
Using this approximation, we can manipulate Eq.~(\ref{eq:om_m1}) to
give constraints on either $\Omega_m$ or $h$
\begin{eqnarray}
  \Omega_m &=& (0.282 \pm 0.018)
  \left(\Omega_m h^2\over 0.1326\right)^{0.58}\nonumber\\
  & & \hspace{1.0cm}
    \times\left[1 + 0.25 \Omega_k + 0.23(1+w)\right], 
    \label{eq:omconstraint}\\
  h &=& (0.686 \mp 0.022)
    \left(\Omega_m h^2\over 0.1326\right)^{0.21}\nonumber\\
  & & \hspace{1.0cm}
    \times\left[1 - 0.13 \Omega_k - 0.12(1+w)\right] 
    \label{eq:hconstraint}.
\end{eqnarray}
The additional uncertainty in $\Omega_m$, $\pm0.018$ in
Eq.~(\ref{eq:omconstraint}) compared with $\pm0.15$ in
Eq.~(\ref{eq:om_m1}), is produced by the dependence of the distance
ratio on $\Omega_m$. In Eqns.~(\ref{eq:omconstraint})
\&~(\ref{eq:hconstraint}), the uncertainty in the first terms are
correlated so as to leave $\Omega_m h^2$ constant. One should
additionally include the errors from $\Omega_m h^2$, $\Omega_k$, and
$w$, although these are consistent between the two results.

Looking at the fractional error in $\Omega_m$, the contribution from
the uncertainty in the SDSS acoustic scale is about 6\%, that from the
uncertainty in $\Omega_m h^2$ is about 2\%, that from $w$ is about 3\%
if the error on $w$ is 10\%, and that from curvature is below 1\%
unless the cosmology is rather non-standard. Hence our result is still
limited by the SDSS-II BAO data volume and not by our knowledge of the
other cosmological parameters in Eq.~(\ref{eq:omconstraint}). Of
course, these expressions only hold for mild perturbations from the
concordance cosmology; for other cases, one should return to the raw
distance constraints. We note that these expressions have not used the
angular acoustic scale in the CMB, so they are independent of what is
happening with dark energy at $z>0.35$.

\begin{figure}
  \centering
  \resizebox{0.98\columnwidth}{!}{\includegraphics{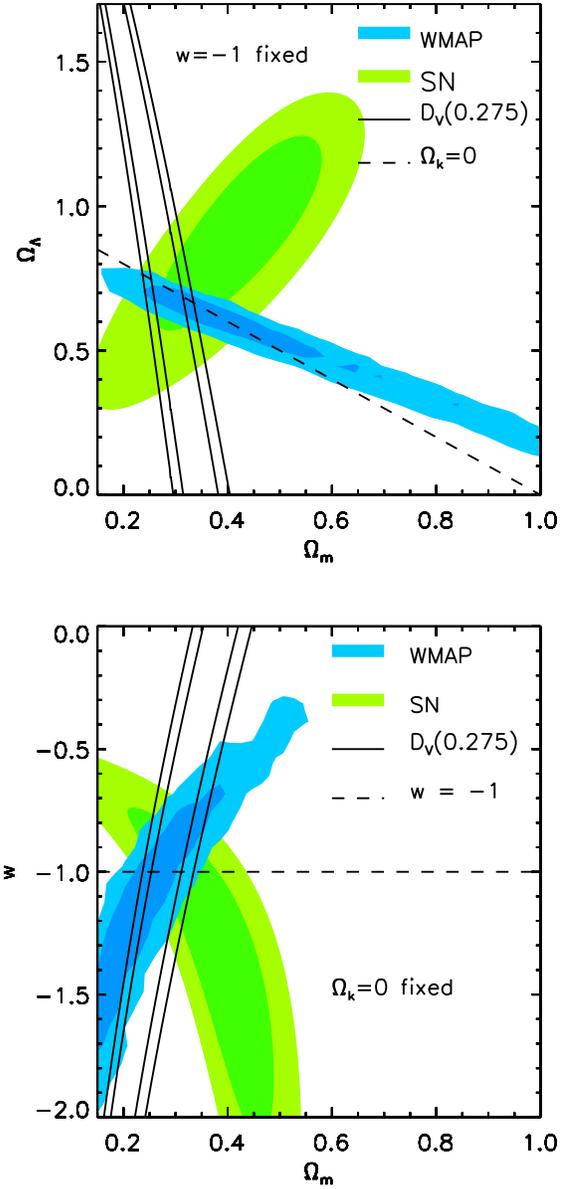}}
  \caption{Cosmological constraints on $\Lambda$CDM cosmologies (upper
    panel) and flat CDM models where we allow $w$ to vary (lower
    panel), from WMAP5 (blue), Union supernova (green) and our
    constraint on $r_s/D_V(0.275)$ (solid contours). Contours are
    plotted for $-2\ln{\cal L}/{\cal L}_{\rm true}<2.3,\,6.0$,
    corresponding to 68\% and 95\% confidence intervals. The dashed
    lines show flat models (upper panel) and $\Lambda$ models (lower
    panel).}
  \label{fig:simple_cosmo}
\end{figure}
Fig.~\ref{fig:simple_cosmo} shows the BAO constraints from
Eq.~(\ref{eq:DV0275}) on $\Omega_m$ and $\Omega_\Lambda$ for
$\Lambda$CDM cosmologies (upper panel), and on $\Omega_m$ and $w$ for
flat models where constant $w\ne-1$ is allowed (lower panel). We take
a Gaussian prior of $\Omega_m h^2 = 0.1326 \pm 0.0063$ and assume that
the error on $\Omega_bh^2$ is negligible as the WMAP5 data already
constrain it to 0.5\% \citep{komatsu09}. These constraints exclude the
angular acoustic scale in the CMB, so they are independent of the dark
energy behaviour at the redshifts beyond our sample. For comparison we
plot the full WMAP5 constraints \citep{komatsu09}, which include the
constraints on the distance to last scattering, and constraints from
the Union supernova sample \citep{kowalski08}, which constrain angular
diameter distance ratios up to $z\sim1$. Results from full likelihood
fits combining these data are presented in
Section~\ref{sec:cosmoparams}.

\section{Testing the Robustness of the Results} \label{sec:robust}

\subsection{The effect of redshift-space distortions}  \label{sec:zspace}

\begin{figure}
  \centering
  \resizebox{0.9\columnwidth}{!}{\includegraphics{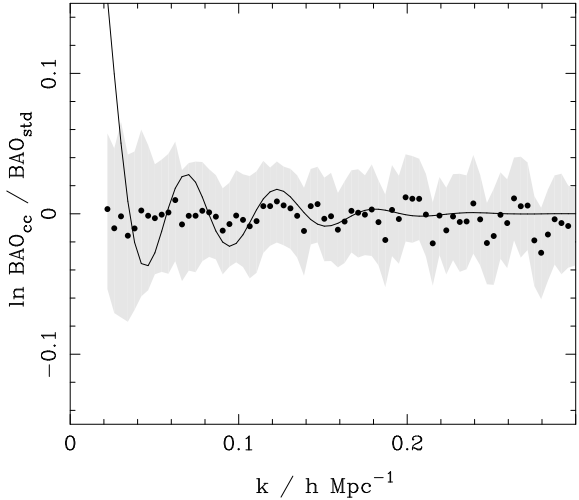}}
  \caption{The log ratio between the BAO recovered from the SDSS LRG
    power spectrum and the power spectrum of the halo catalogue
    derived from the LRG sample as described by \citet{reid09} (solid
    circles). For comparison we plot the BAO expected for a flat
    $\Lambda$CDM model with $\Omega_m=0.25$, $h=0.72$, \&
    $\Omega_bh^2=0.0223$ (solid line), and the errors on each
    measurement (grey shaded region). There are no oscillatory
    features induced by the cluster-collapse procedure, and the
    scatter is well within the errors.}
  \label{fig:bao_cmpr_cc}
\end{figure}
We have fitted our spline $\times$ BAO model to the observed SDSS LRG
power spectrum, as calculated by \citet{reid09}, where the galaxy
power spectrum and derived cosmological constraints are
presented. Using numerical simulations, a scheme is presented in
\citet{reid09} to recover the halo power spectrum from the LRG
distribution by only keeping a single LRG within each halo. We have
fitted both the galaxy and the halo power spectra with our spline
$\times$ BAO model. The log ratio between the BAO recovered in the
resulting fits is shown in Fig.~\ref{fig:bao_cmpr_cc}. This shows that
the cluster-collapse correction for these galaxies results in a smooth
change in the power spectrum on the scales fitted, and does not alter
the position or amplitude of the BAO in a significant way.

Because of the different galaxy properties within the SDSS main galaxy
sample, and the 2dFGRS, we do not attempt to correct for the more
complicated distribution of galaxies within the haloes of that sample,
and recover the halo power spectrum. In contrast, the halo occupation
distribution of the SDSS LRGs is simple, in that there is only a
single population of galaxies that are predominantly central rather
than satellite galaxies in their hosting haloes \citep{reid09}.  But
we have seen that for LRGs, the correction is smooth, and we expect
this to be true for the galaxies at $z<0.2$ as well.

\subsection{Sample selection}  \label{sec:sample}

\begin{figure*}
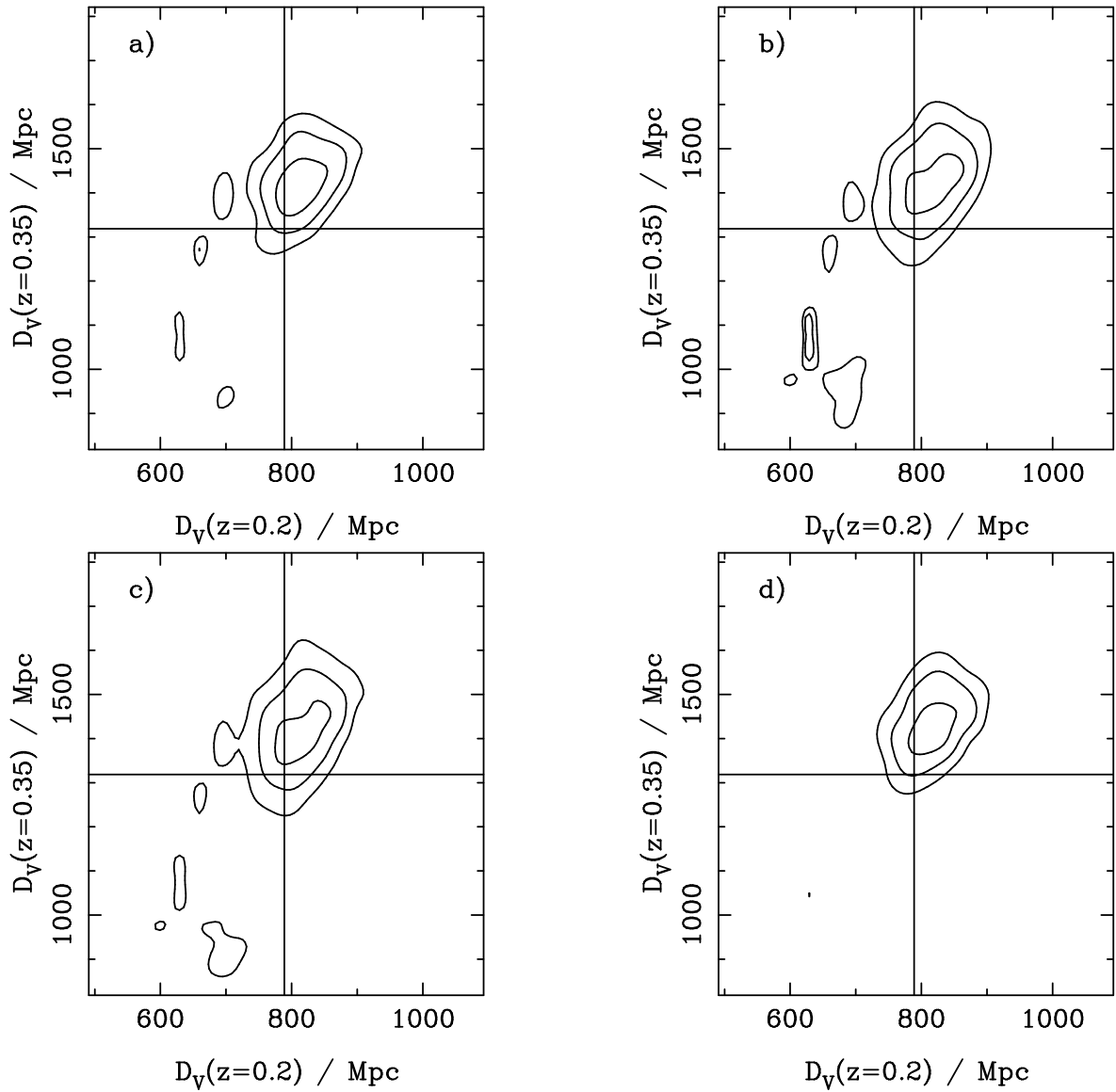

  \begin{center}
  \resizebox{0.8\columnwidth}{!}{\includegraphics{like_a.ps}}
  \hspace{2cm}
  \resizebox{0.8\columnwidth}{!}{\includegraphics{like_b.ps}}\\
  \resizebox{0.8\columnwidth}{!}{\includegraphics{like_c.ps}}
  \hspace{2cm}
  \resizebox{0.8\columnwidth}{!}{\includegraphics{like_d.ps}}\\
  \end{center}
  \caption{As Fig.~\ref{fig:like_data3}, but now considering results
    from 4 choices of catalogue: (a) all SDSS and non-overlapping
    2dFGRS data, (b) excluding both the early SDSS data and 2dFGRS,
    (c) excluding the 2dFGRS, (d) excluding the early SDSS
    data.}
\label{fig:like_data1a}
\end{figure*}

\begin{table}
\begin{center}
\begin{tabular}{lllll}
  \hline
      & $d_{0.2}$            & $d_{0.35}$ \\
  \hline
  (a) default               & $0.1905\pm0.0061$ & $0.1097\pm0.0036$ \\
  (b) no early SDSS, 2dFGRS & $0.1923\pm0.0072$ & $0.1102\pm0.0041$ \\
  (c) no 2dFGRS             & $0.1907\pm0.0062$ & $0.1090\pm0.0036$ \\
  (d) no early SDSS         & $0.1917\pm0.0069$ & $0.1109\pm0.0044$ \\
  (e) fit to three $P(k)$   & $0.1901\pm0.0066$ & $0.1080\pm0.0043$ \\
  (f) original calibration  & $0.1919\pm0.0071$ & $0.1094\pm0.0046$ \\
  (g) varying $D_{\rm damp}$ & $0.1918\pm0.0080$ & $0.1100\pm0.0048$ \\
  (h) $\langle n(z)\rangle$ sampling galaxies & $0.1890\pm0.0068$ & $0.1102\pm0.0045$ \\
  \hline
\end{tabular}
\caption{\label{table:bao_DV_constraints} Measurements of $d_z\equiv
  r_s(z_d)/D_V(z)$ at $z=0.2$ and $z=0.35$ from the different analysis runs
  described in the captions to Figs.~\ref{fig:like_data1a}
  \&~\ref{fig:like_data1b}.}
\end{center}
\end{table}

We have run our full analysis pipeline using three subsamples of
galaxies.  Results from fits to $D_V(z)$ with two nodes are shown in
Fig.~\ref{fig:like_data1a}, for different catalogues, given
$r_s(z_d)=154.7\,{\rm Mpc}$. The best-fit constraints for these models
on $d_z$ are given in Table~\ref{table:bao_DV_constraints}. Our
default analysis is included in panel (a) for comparison. Here, we
analyse data from the SDSS and the 2dFGRS, including the early SDSS
data, where we cut the sample at the extinction-corrected magnitude
limit $r<17.5$. We compare with results obtained (b) excluding the
early SDSS data and the 2dFGRS, (c) using just the SDSS data, and (d)
excluding the early SDSS data but including the 2dFGRS. Including the
early SDSS galaxies decreases the errors at redshift $z=0.2$ and
$z=0.35$ by approximately 14\%. Including the 2dFGRS galaxies has a
smaller effect, decreasing the error at $z=0.2$ by approximately
4\%. The parameters of the best-fit solutions do not move
significantly with any of the sample changes: $d_{0.2}$ moves by a
maximum of 0.3$\sigma$, while $d_{0.35}$ moves by a maximum of
0.2$\sigma$. The inclusion of the 2dFGRS actually moves the best-fit
solution for $D_V(0.35)/D_V(0.2)$ slightly towards that of a
concordance $\Lambda$CDM model.

\subsection{The number of redshift slices included}

\begin{figure*}
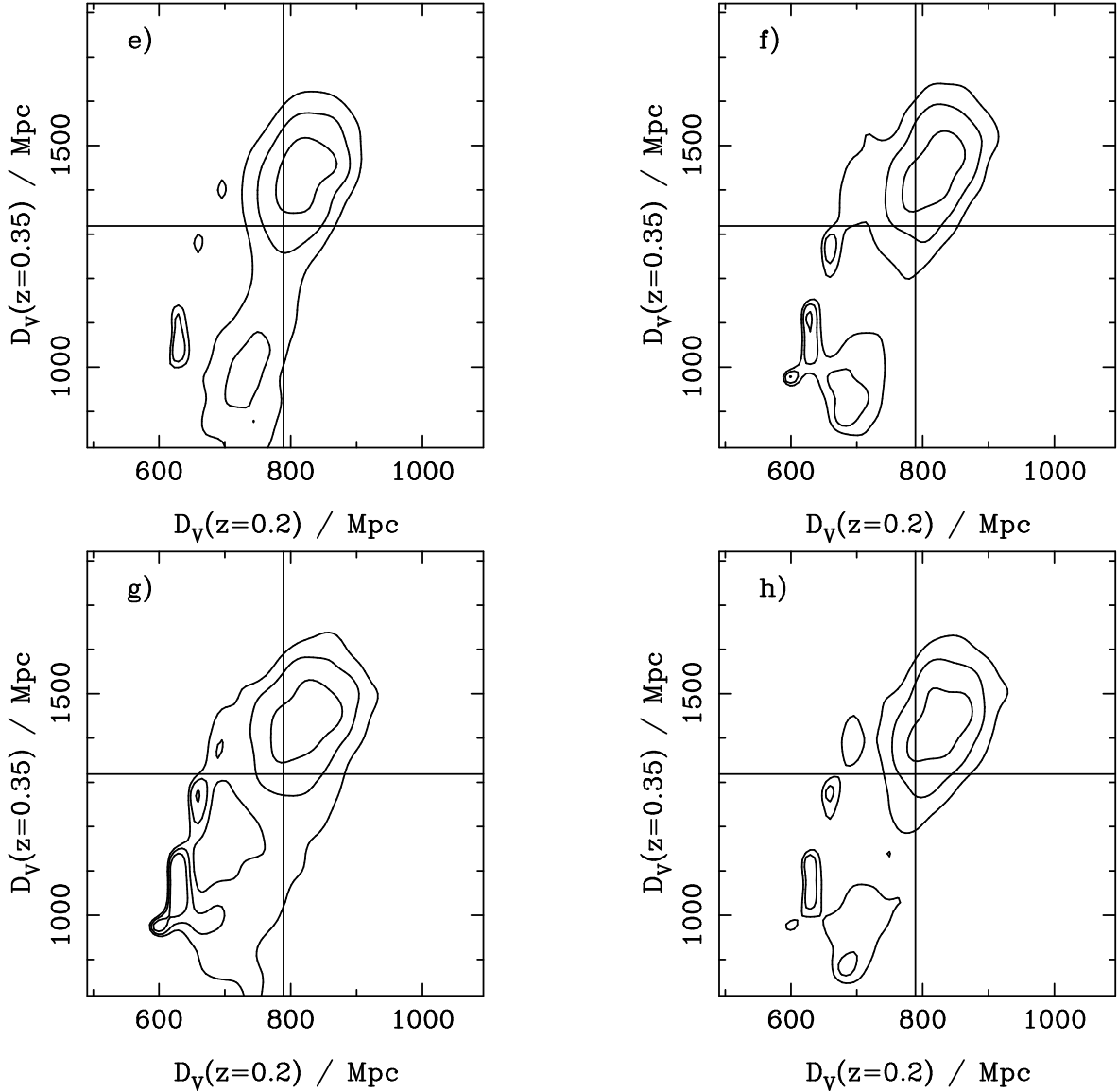

  \begin{center}
  \resizebox{0.8\columnwidth}{!}{\includegraphics{like_e.ps}}
  \hspace{2cm}
  \resizebox{0.8\columnwidth}{!}{\includegraphics{like_f.ps}}\\
  \resizebox{0.8\columnwidth}{!}{\includegraphics{like_g.ps}}
  \hspace{2cm}
  \resizebox{0.8\columnwidth}{!}{\includegraphics{like_h.ps}}\\
  \end{center}
  \caption{Likelihood contour plots as in Fig.~\ref{fig:like_data3},
    for the SDSS data, excluding the early data and the 2dFGRS, but
    now calculated for (e) fit to three power spectra, (f) old rather
    than uber-calibration. (g) allowing the BAO damping scale to vary
    with a simple Gaussian prior $D_{\rm damp}=10\pm5\mpcoh$, (h)
    randomized galaxy redshifts used to give the expected radial
    galaxy distribution.}
\label{fig:like_data1b}
\end{figure*}
We now consider the robustness of our fit to the number of redshift
slices analysed. This test was performed on the conservative data
sample, excluding the early SDSS data and the 2dFGRS. In our default
analysis we fit power spectra calculated for six redshift slices, and
the resulting likelihood surface for the late SDSS sample is shown in
panel (b) of Fig.~\ref{fig:like_data1a}. For comparison, panel (e) of
Fig.~\ref{fig:like_data1b} shows the likelihood surface calculated
using power spectra from only three redshift slices (details of the
slices chosen are presented in Section~\ref{sec:slices}). Because we
are only fitting two $D_V(z)$ nodes, these should be constrained by
our reduced fit using three redshift slices. Panel (e) of
Fig.~\ref{fig:like_data1b} shows that this is true, but comparison
with panel (b) of Fig.~\ref{fig:like_data1a} shows that the
constraints are tighter if we model power spectra from six redshift
slices. Clearly, extra information is available from the extra
redshift slices, and we therefore fit to six redshift slices for our
default analysis.

\begin{figure*}
\centering
\resizebox{0.9\textwidth}{!}{\includegraphics{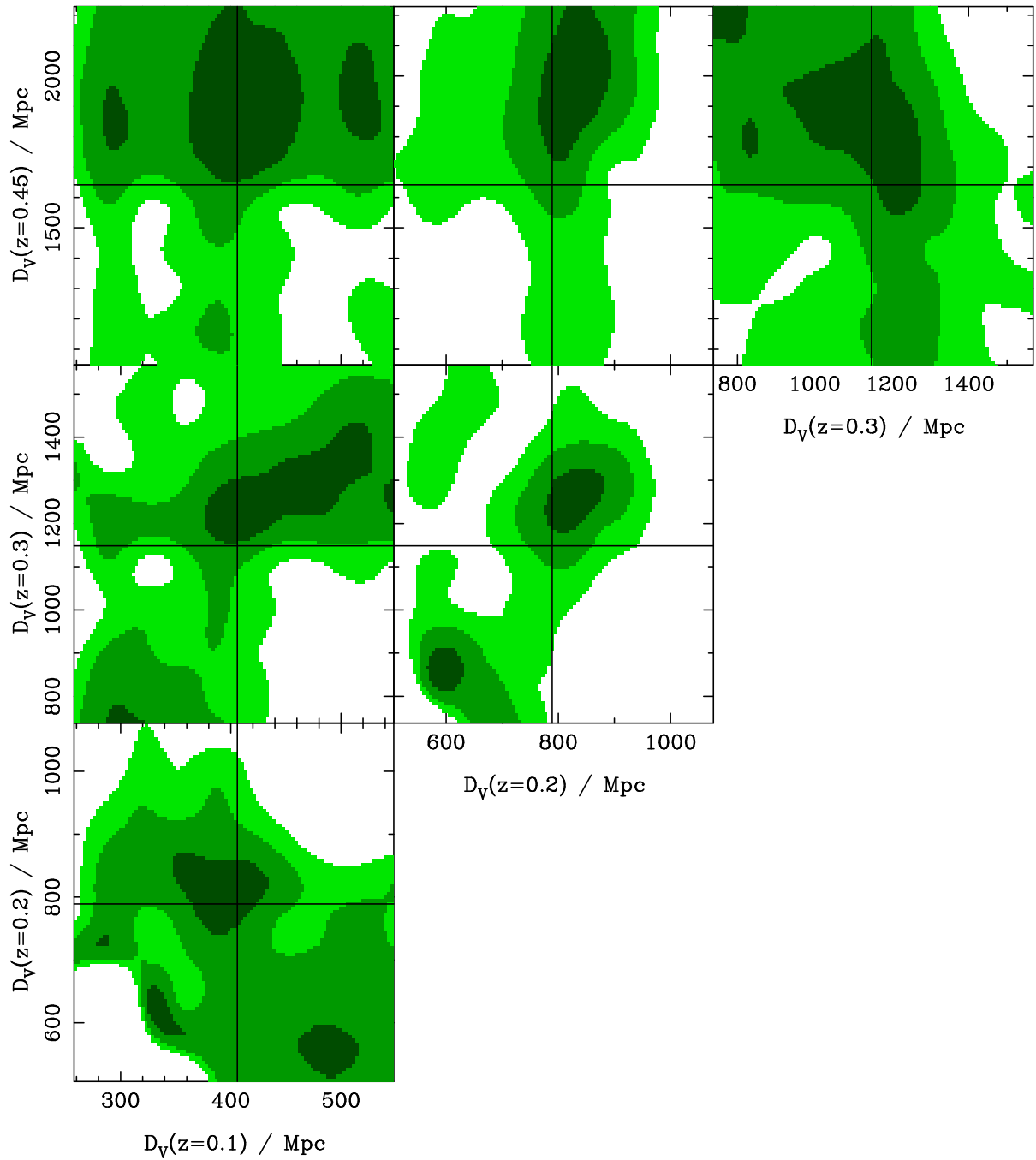}}
\caption{Contour plots showing slices through the likelihood for four
  $D_V(z)$ cubic spline nodes at $z=0.1$, $z=0.2$ $z=0.3$ and
  $z=0.45$, calculated for our default analysis using six power
  spectra, uber-calibration, and a fixed BAO damping scale of $D_{\rm
    damp}=10\mpcoh$. Shaded regions are plotted for $-2\ln{\cal
    L}/{\cal L}_{\rm true}<2.3,\,6.0,\,9.3$, which for a multi-variate
  Gaussian distribution with two degrees of freedom correspond to
  68\%, 95\% and 99\% confidence intervals. Likelihoods were adjusted
  to match these Gaussian confidence intervals as described in
  Section~\ref{sec:bao_ln}. In each panel, the nodes that are not
  shown were fixed at the default $\Lambda$CDM ($\Omega_m=0.25$,
  $\Omega_\Lambda=0.75$) values. We use shaded regions in this plot to
  show the likelihood surface, compared with the contours in
  Figs.~\ref{fig:like_data1a} \&~\ref{fig:like_data1b} because the
  likelihood surface is more complicated with four nodes, and the
  shading helps to distinguish peaks from troughs.}
\label{fig:like_data2}
\end{figure*}
It is interesting to test if there is sufficient information to
constrain the shape of $D_V(z)$ beyond our simple spline model with 2
nodes. Results from fits allowing four $D_V(z)$ nodes are shown in
Fig.~\ref{fig:like_data2}. There is a clear maximum in the slices
through the likelihood surface close to the $\Lambda$CDM model, but
the surface is noisy, and there are secondary maxima present. There is
a strong degeneracy between $D_V(0.3)$ and $D_V(0.45)$, and between
$D_V(0.1)$ and $D_V(0.2)$: the data contain limited information to
distinguish the shape of the distance-redshift relation between these
redshifts. Consequently, we do not try to extract this information,
instead concentrating on fits where there are only two nodes in
$D_V(z)$.

\subsection{The covariance matrix}

Because we are analysing overlapping shells in redshift, the power
spectra will be strongly correlated and the estimation of the
covariance matrix will be in error if we do not have sufficient mock
catalogues. In order to test this, we have recalculated our covariance
matrix using $1/3$ as many LN catalogues, and have used this matrix to
recalculate the required corrections to the confidence intervals using
independent sets of LN catalogues. We find consistent results in the
factors required to match the confidence intervals to those expected
for a multi-variate Gaussian distribution. We have also performed a
full analysis using this reduced covariance matrix, and find results
consistent with using our default covariance matrix.

\subsection{Calibration}

The likelihood surface shown in panel (f) of
Fig.~\ref{fig:like_data1b} was calculated using a SDSS galaxy sample
with luminosities calibrated using the photometric calibration
\citep{tucker06}, prior to the uber-calibration analysis
\citep{padmanabhan08}. This affects the calculation of the redshift
completeness for any region observed, and also the
luminosity-dependent weights applied to the SDSS galaxies. The effect
of this calibration change on our results is small, and there is no
significant change between the likelihood surface in panel (f) of
Fig.~\ref{fig:like_data1b} and that in panel (b) of
Fig.~\ref{fig:like_data1a}, where the uber-calibration data set was
used.

\subsection{BAO damping scale}  \label{sec:Ddamp}

Panel (g) of Fig.~\ref{fig:like_data1b} shows the likelihood surface
if we allow the BAO damping scale to be a free parameter in the fit,
placing a simple Gaussian prior on its value $D_{\rm
  damp}=10\pm5\mpcoh$. This prior on the BAO damping scale is
conservative. From simulations, \citet{reid08} found $D_{\rm
  damp}=9.2\pm1\mpcoh$, with no variation with redshift for $0<z<0.5$
for halo density fields, and $D_{\rm damp}=9.7\pm1\mpcoh$ for density
fields matched to the LRGs. The mild cosmological dependence suggested
by \citet{eisenstein07} shows that the main cosmological dependence is
through the linear growth rate; current constraints on $\sigma_8$ are
much better than that required to significantly change $D_{\rm damp}$,
and we consider $\pm 5\mpcoh$ to be a conservative prior. Allowing the
damping scale to vary degrades the constraint, increasing the size of
the parameter confidence regions. The best-fit solution does not move
significantly, suggesting that our default assumption of a fixed
damping scale is sufficiently accurate to current data precision.

\subsection{Radial galaxy distribution model}  \label{sec:zfit}

Finally, analysis run (h) shows the constraints if we use a random
catalogue where we randomly choose a galaxy redshift for each angular
position chosen. i.e. to model the expected redshift distribution
$\langle n(z)\rangle$, we sample from the galaxy redshift
distribution. This test was designed to investigate the dependence of
the analysis on how well we model the radial galaxy
distribution. Randomly sampling galaxies to obtain this distribution,
perfectly matches the redshift distribution of the galaxies and that
of the random catalogue used to define the survey region. In fact, we
see no change in our results if we do this rather than using a smooth
fit to the redshift distribution. This gives us confidence that our
results are not sensitive to this modelling.

\subsection{Secondary likelihood maxima}  \label{sec:secondaries}

In the likelihood surfaces in Figs.~\ref{fig:like_data1a}
\&~\ref{fig:like_data1b}, we see secondary likelihood maxima, which
appear to lie on a degeneracy stretching from $D_V(0.2)=700$\,Mpc,
$D_V(0.35)=1500$\,Mpc to $D_V(0.2)=600$\,Mpc,
$D_V(0.35)=1000$\,Mpc. These minor peaks in the likelihood, which
appear as isolated islands in the likelihood surface are of lower
significance than the strong peak close to the parameters of a
concordance $\Lambda$CDM model. Tests have shown that the secondary
peaks result from the interplay of two competing effects, which are
themselves a result of using the wrong cosmology to analyse the
BAO. These are:
\begin{enumerate}
\item A shift in the BAO position,
\item An increase in the width of the window associated with each
  band-power, caused by BAO in different redshift shells being
  out-of-phase. This can smooth out the BAO signal.
\end{enumerate}
Secondary maxima are produced where the BAO shift and the smoothing
``balance''. If we redo the analysis ignoring the second effect by
assuming that the window function is a $\delta$-function centred on
the peak, these secondary maxima are removed.

\subsection{Dependency on $D_V$}  \label{sec:D_V_test}

\begin{figure}
\centering
\resizebox{0.9\columnwidth}{!}{\includegraphics{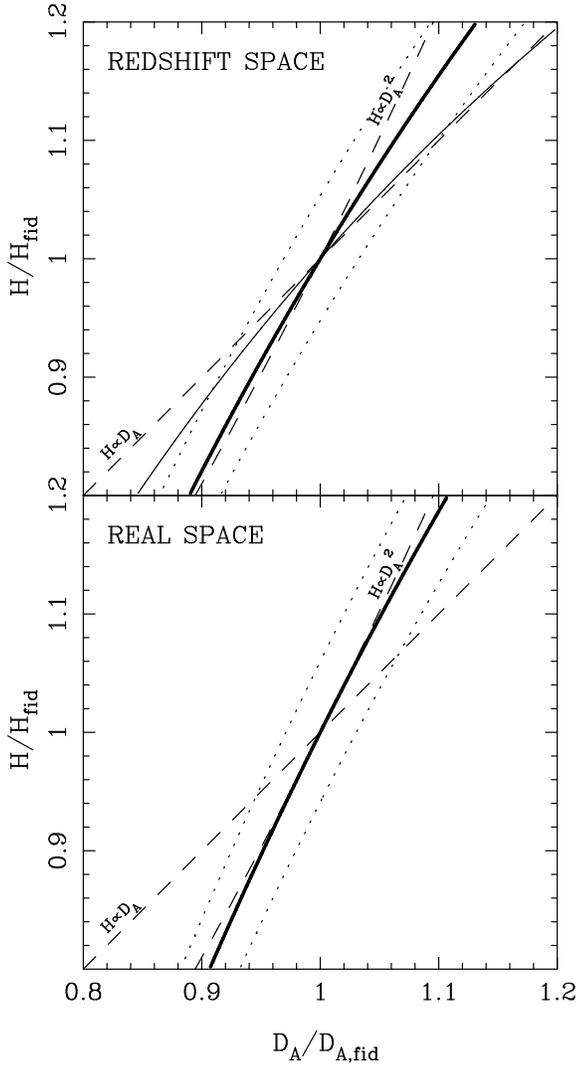}}
\caption{The expected shift recovered from an analysis of the BAO
  position in a spherically averaged galaxy power spectrum, if there
  are radial and angular distortions induced by assuming an incorrect
  cosmology when analysing the data. The thick solid contour shows no
  residual shift, while the dotted contours show a 1\% shift. For
  comparison we plot the expected behaviour for an isotropic power
  spectrum $H(z)\propto D_A^{2}$, and for an increased importance of
  the radial distortion $H(z)\propto D_A$ (dashed lines). The top
  panel approximates redshift-space, by weighting the power in the
  spherical average by $(1+\beta\mu^2)^2$, with $\beta=0.25$, matching
  that expected for the SDSS LRGs, while the bottom panel does not
  include this weighting. For comparison, the thin solid contour in
  the top panel marks no residual shift for data with $\beta=1$,
  showing that we should expect the radial signal to increase in
  importance for such a sample.}
\label{fig:D_V_test}
\end{figure}

A possible concern about our method of analysis is that we assume a
fiducial $\Lambda$CDM model to convert redshifts to comoving
coordinate distances and measure the position of the BAO in the
spherically-averaged power spectrum. If the true cosmological model
has different angular diameter distance-redshift relation $D_A(z)$ and
Hubble parameter $H(z)$ than this fiducial model, this would cause
angular and radial distortions in the density field from which we
estimate the power spectrum. By presenting results in terms of $D_V$
we remove the anisotropic information, and assume that the expected
BAO position for all cosmological models is solely dependent on their
predicted value of $D_V$. This must break down for models that behave
very differently from our fiducial $\Lambda$CDM model.

We now test the sensitivity of the assumption that the BAO position in
the spherically averaged power spectrum only depends on $D_V$ for
cosmological models that predict significant anisotropic distortions
in the density field away from our fiducial model. To do this, we
compute the shifts of the BAO position expected when one measures the
spherically-averaged power in either real or redshift space for such
models. To simplify the analysis, we assume that the BAO in the
spherically averaged $P(k)$ will be shifted by the average of the
shifts in ${\bf k}$ predicted over all angles: ie. our BAO fit
recovers the weighted mean shift in the 3D power. In redshift-space we
also follow the distant observer approximation, and assume that the
angular dependence of the true 3D power spectrum is given by
$(1+\beta\mu^2)^2$, where $\mu$ is the cosine of the angle to the
line-of-sight and $\beta=\Omega_m^{0.55}/b$. The anisotropy in the
observed power spectrum caused by redshift-space distortions will act
as a weight when we spherically average.

For the SDSS LRGs, which provide most of our cosmological signal, we
take an effective redshift of $z=0.35$, and assume a $\Lambda$CDM
model with $\Omega_m(z=0)=0.25$, giving $\Omega_m(z=0.35)=0.45$. The
LRGs are strongly biased and the model of \citet{tegmark04} gives an
effective relative bias for our sample, which we correct for in the
power spectrum calculation, of $\langle b/b_*\rangle=1.9$. Matching
the normalisation of the measured LRG power spectrum \citep{reid09}
gives that $b_*=1.34$ assuming the LRG clustering is constant in
comoving coordinates (e.g. \citealt{percival07b}), and that
$\sigma_8({\rm matter},z=0)=0.8$, so $\sigma_8({\rm
  matter},z=0.35)=0.68$ \citep{komatsu09}.  This suggests that we
should expect $\beta\sim0.25$ for the LRG power spectrum, and we show
contours calculated assuming $\beta=0.25$ in Fig.~\ref{fig:D_V_test},
which we compare with the prediction for $\beta=1$. Note that our
luminosity-dependent weighting means that we are upweighting highly
biased galaxies, and that our analysis will therefore have a smaller
effective $\beta$ than analyses without such weighting, such as the
measurements presented by \citet{cabre09}. 

Fig.~\ref{fig:D_V_test} shows the relation between radial and angular
distortions, $H/H_{\rm fid}$ and $D_A/D_{A,{\rm fid}}$, which give
rise to zero and $\pm 1\%$ shift in the spherical averaged power
spectrum. Here, $H_{\rm fid}$ is the fiducial value of $H$, and
similarly for $D_A$. For general cosmological models, $H/H_{\rm fid}$
and $D_A/D_{A,{\rm fid}}$ will depend on redshift, so that the final
effective shift will be an average over a trajectory in this diagram
which is determined by the model to be tested. Fig.~\ref{fig:D_V_test}
also shows the expected line of zero average shift we would expect if
the BAO position only depends on $D_V(z)$, which would lead to
behaviour such that $H(z)\propto D_A^{2}$. For comparison, we show the
prediction for a model with increased importance of the radial
distortions, with $H(z)\propto D_A$. This is included because we would
expect that the redshift-space distortions will increase the
importance of the radial information. However, the $H(z)\propto
D_A^{2}$ line is a significantly better fit, even in
redshift-space. The $H(z)\propto D_A^{2}$ line does not cross the
contours marking a 1\% average shift for our redshift-space power
spectrum, showing that the assumption that the recovered BAO position
only depends on $D_V$ at most produces a 1\% systematic in the
best-fit for models with an anisotropy distortion away from our
fiducial model of up to 20\% in the radial direction. Such a 1\%
systematic shift, which requires a model that is extremely discrepant
from $\Lambda$CDM, is significantly below the statistical precision of
our 2.7\% accuracy distance measurement. It is therefore a reasonable
approximation to use our measurements of $D_V$ to constrain a wide
variety of cosmological models.

\section{Cosmological Parameter Constraints} \label{sec:cosmoparams}

We now apply our full constraints to a cosmological parameter
analysis. We assume that the likelihood of a model is given by a
multi-variate Gaussian distribution around the $D_V(z)$ measurements
given by Eq.~(\ref{eq:DV_constraints}), with the inverse covariance
matrix of Eq.~(\ref{eq:DV_invcov}). Throughout this section we
consider four models: a flat universe with a cosmological constant
($\Lambda$CDM), a $\Lambda$CDM universe with curvature
(o$\Lambda$CDM), a flat universe with a dark energy component with
constant equation of state $w$ (wCDM), and a wCDM universe with
curvature (owCDM). This is the same model set considered by
\citet{reid09}. We use a modified version of {\sc cosmomc}
\citep{lewis02} to perform the likelihood calculations.

\subsection{SN + BAO + CMB prior likelihood fits}  \label{sec:CMBprior}

\begin{table*}
\begin{center}
\begin{tabular}{lllll}
  \hline
  parameter & $\Lambda$CDM & o$\Lambda$CDM & wCDM &
  owCDM \\
  \hline
$\Omega_m$ & 0.288 $\pm$ 0.018 & 0.286 $\pm$ 0.018 & $0.290^{+0.018}_{-0.019}$ & 0.286 $\pm$ 0.018  \\
$H_0$ & $68.1^{+2.2}_{-2.1}$ & 68.6 $\pm$ 2.2 & 67.8 $\pm$ 2.2 & 68.2 $\pm$ 2.2  \\
$\Omega_k$ &  - & -0.097 $\pm$ 0.081 &  - & $-0.199^{+0.080}_{-0.089}$  \\
$w$ &  - &  - & -0.97 $\pm$ 0.11 & $-0.838^{+0.083}_{-0.084}$  \\
$\Omega_\Lambda$ & 0.712 $\pm$ 0.018 & $0.811^{+0.084}_{-0.085}$ & $0.710^{+0.019}_{-0.018}$ & $0.913^{+0.092}_{-0.082}$  \\
\hline
$d_{0.275}$& 0.1381 $\pm$ 0.0034 & 0.1367 $\pm$ 0.0036 & 0.1384 $\pm$ 0.0037 & 0.1386 $\pm$ 0.0037 \\
$D_V(0.275)$& 1111 $\pm$ 31 & 1120 $\pm$ 33 & 1109 $\pm$ 32 & $1108^{+32}_{-33}$ \\
$f$& $1.662 \pm 0.004$ & $1.675 \pm 0.011$ & $1.659 \pm 0.011$ & 1.665 $\pm$ 0.011 \\	
Age (Gyr) & $14.02^{+0.32}_{-0.31}$ & 14.43 $\pm$ 0.48 & 13.95 $\pm$ 0.36 & 14.38 $\pm$ 0.44  \\
\hline
\end{tabular}
\caption{\label{table:baoSNandCMBprior} Marginalized one-dimensional
  constraints (68\%) for BAO$+$SN for flat $\Lambda$CDM, $\Lambda$CDM
  with curvature (o$\Lambda$CDM), flat wCDM (wCDM), and wCDM with
  curvature (owCDM). The non-standard cosmological parameters are
  $d_{0.275}\equiv r_s(z_d)/D_V(0.275)$ and $f\equiv
  D_V(0.35)/D_V(0.2)$. We have assumed priors of $\Omega_c h^2 =
  0.1099 \pm 0.0063$ and $\Omega_b h^2 = 0.02273 \pm 0.00061$,
  consistent with WMAP5-only fits to all of the models considered
  here.  We also impose weak flat priors of $-0.3 < \Omega_k < 0.3$
  and $-3 < w < 0$.}
\end{center}
\end{table*}

We first consider the constraints excluding the angular acoustic scale
in the CMB, in order to consider data that are independent of the dark
energy behaviour at the redshifts beyond our sample. This is important
because it ensures that our results only depend on the acceleration of
the Universe at late times and so do not depend on so-called early
dark energy models \citep{wetterich88,ratra88,zlatev99,steinhardt99},
which have non-negligible dark energy at early times. We take Gaussian
priors $\Omega_c h^2 = 0.1099 \pm 0.0063$ and $\Omega_b h^2 = 0.02273
\pm 0.00061$ from the CMB; these constraints from the ratio of peak
heights in the WMAP5 data alone do not relax when $\Omega_k$ and $w$
are allowed to vary. We also impose weak priors on $-0.3 < \Omega_k <
0.3$ and $-3 < w < 0$.  The parameter constraints from the combination
of Union supernova (SN) \citep{kowalski08} and BAO likelihoods with
these priors are presented in Table~\ref{table:baoSNandCMBprior}.  The
best-fit value of $\Omega_m$ ranges from 0.286 to 0.290, with the 68\%
confidence interval, $\pm 0.018$, while the mean value of $H_0$ varies
between $67.8\Hunit$ and $68.6\Hunit$, and the 68\% confidence
interval remains $\pm 2.2\Hunit$ throughout the four models. In
Section~\ref{sec:interp} we derived BAO only constraints of $\pm
0.018$ on $\Omega_m$ and $\pm 2.2\Hunit$ on $H_0$, for fixed
$\Omega_mh^2$. If we include the 4.8\% error on $\Omega_mh^2$ from the
WMAP5 measurement, then we should expect these errors to increase to
$\pm 0.019$ on $\Omega_m$ and $\pm 2.3\Hunit$ on $H_0$. These agree
perfectly with the {\sc COSMOMC} results if we exclude the supernova
data, so the small difference between the errors in
Table~\ref{table:baoSNandCMBprior} and those expected is caused by the
supernova data helping to constrain $\Omega_m$ and $H_0$
slightly. Similarly, the best-fit values of these parameters agree for
{\sc COSMOMC} results excluding the supernova data. Comparison between
Table~\ref{table:baoSNandCMBprior} and Section~\ref{sec:interp} shows
that the inclusion of the supernova data is moving the best-fit
slightly: $+0.004$ in $\Omega_m$ and $-0.5$ in $H_0$ for the
$\Lambda$CDM model. The {\sc COSMOMC} analysis therefore validates the
simple derivation presented in Section~\ref{sec:interp}. In the space
of models considered here, the BAO constraint on $D_V(0.275)$ already
restricts $D_V(0.35)/D_V(0.2)$ to a much smaller region than our
constraint in Eq.~(\ref{eq:frat}) allows.  While the combination of
these data and our priors are unable to constrain $\Omega_k$, $w$ is
constrained at the $\pm 0.11$ level. For the owCDM model, the weak
prior on $\Omega_k$ leads to an apparent constraint on $w$, but these
errors depend strongly on the prior.

\begin{figure}
  \centering
  \resizebox{0.9\columnwidth}{!}{\includegraphics{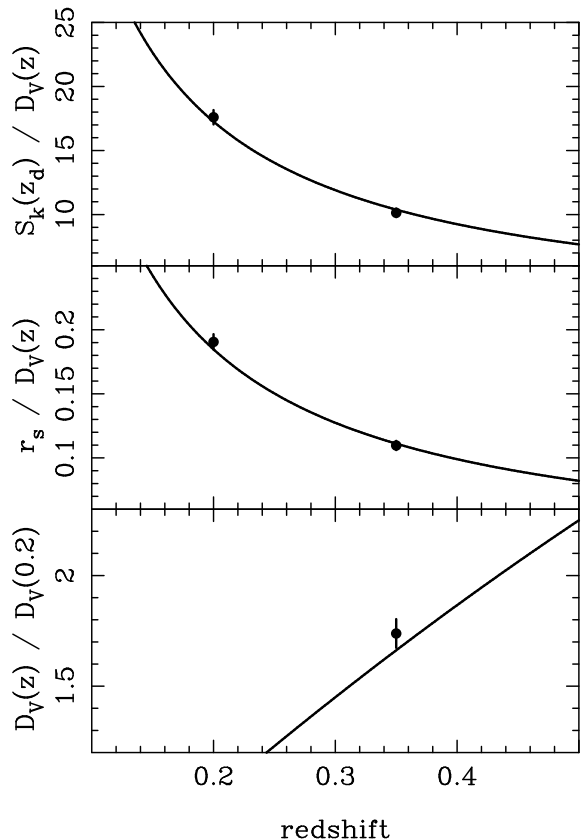}}
  \caption{The BAO constraints (solid circles with 1$\sigma$ errors),
    compared with the best-fit $\Lambda$CDM model. The three panels
    show different methods of using the data to constrain models.}
  \label{fig:DVdata}
\end{figure}
The data are compared with the best-fit $\Lambda$CDM model in
Fig.~\ref{fig:DVdata}. Three ways of considering the data constraints
are shown in different panels. In the bottom panel we plot
$D_V(z)/D_V(0.2)$, which corresponds to matching the geometry at
$z=0.2$ and $z=0.35$ so the BAO match at these redshifts, without
including information about the comoving position of the BAO. In the
middle panel we plot $r_s(z_d)/D_V(z)$, where we now have to model the
comoving sound horizon at the drag epoch. In the top panel we include
a constraint on the sound horizon projected at the last-scattering
surface as observed in the CMB. Marginalising over the set of flat
$\Lambda$CDM models constrained only by the WMAP5 data gives
$r_s(z_d)/S_k(z_d)=0.010824\pm0.000023$, where $S_k(z_d)$ is the
proper distance to the baryon-drag redshift $z_d=1020.5$, as measured
by WMAP5 team \citep{komatsu09}. Ignoring the negligible error on this
quantity, we combine with the BAO results to measure
$S_k(z_d)/D_V(z)$. This effectively removes the dependence on the
comoving sound horizon at the drag epoch, anchoring the BAO
measurements at high redshift: here we have done this at the
baryon-drag epoch so the CMB constraint has matched sound horizon and
projection distance.

\subsection{CMB + BAO likelihood fits}

\begin{table*}
\begin{center}
\begin{tabular}{llllllll}
  \hline
  parameter & $\Lambda$CDM & o$\Lambda$CDM & wCDM & 
  owCDM & owCDM$+$SN & owCDM$+H_0$ & owCDM$+$SN$+H_0$\\
  \hline
$\Omega_m$ & 0.278 $\pm$ 0.018 & 0.283 $\pm$ 0.019 & 0.283 $\pm$ 0.026 & $0.240^{+0.044}_{-0.043}$ & 0.290 $\pm$ 0.019 & $0.240^{+0.025}_{-0.024}$ & 0.279 $\pm$ 0.016  \\
$H_0$ & 70.1 $\pm$ 1.5 & $68.3^{+2.2}_{-2.1}$ & 69.3 $\pm$ 3.9 & 75.3 $\pm$ 7.1 & 67.6 $\pm$ 2.2 & 74.8 $\pm$ 3.6 & 69.5 $\pm$ 2.0  \\
$\Omega_k$ &  - & $-0.007^{+0.006}_{-0.007}$ &  - & -0.013 $\pm$ 0.007 & -0.006 $\pm$ 0.008 & -0.014 $\pm$ 0.007 & -0.003 $\pm$ 0.007  \\
$w$ &  - &  - & -0.97 $\pm$ 0.17 & $-1.53^{+0.51}_{-0.50}$ & -0.97 $\pm$ 0.10 & $-1.49^{+0.32}_{-0.31}$ & -1.00 $\pm$ 0.10  \\
$\Omega_\Lambda$ & 0.722 $\pm$ 0.018 & 0.724 $\pm$ 0.019 & 0.717 $\pm$ 0.026 & 0.772 $\pm$ 0.048 & 0.716 $\pm$ 0.019 & 0.773 $\pm$ 0.029 & 0.724 $\pm$ 0.018  \\
$100 \Omega_b h^2$ & 2.267 $\pm$ 0.058 & 2.269 $\pm$ 0.060 & 2.275 $\pm$ 0.061 & $2.254^{+0.062}_{-0.061}$ & 2.271 $\pm$ 0.061 & $2.254^{+0.061}_{-0.062}$ & 2.284 $\pm$ 0.061  \\
$\tau$ & 0.086 $\pm$ 0.016 & 0.089 $\pm$ 0.017 & 0.087 $\pm$ 0.017 & 0.088 $\pm$ 0.017 & 0.089 $\pm$ 0.017 & 0.088 $\pm$ 0.017 & $0.089^{+0.017}_{-0.018}$  \\
$n_s$ & 0.961 $\pm$ 0.013 & 0.963 $\pm$ 0.014 & 0.963 $\pm$ 0.015 & 0.958 $\pm$ 0.014 & 0.963 $\pm$ 0.014 & 0.957 $\pm$ 0.014 & 0.964 $\pm$ 0.014  \\
$\ln(10^{10} A_{05})$ & $3.074^{+0.040}_{-0.039}$ & 3.060 $\pm$ 0.042 & 3.070 $\pm$ 0.041 & $3.062^{+0.042}_{-0.043}$ & $3.062^{+0.041}_{-0.042}$ & 3.062 $\pm$ 0.042 & 3.072 $\pm$ 0.042  \\
\hline
$d_{0.275}$& 0.1411 $\pm$ 0.0030 & 0.1387 $\pm$ 0.0036 & $0.1404^{+0.0036}_{-0.0035}$ & 0.1382 $\pm$ 0.0037 & 0.1379 $\pm$ 0.0036 & $0.1387^{+0.0036}_{-0.0037}$ & $0.1402^{+0.0033}_{-0.0034}$\\
$D_V(0.275)$& 1080 $\pm$ 18 & $1110^{+32}_{-31}$ & 1089 $\pm$ 31 & 1111 $\pm$ 33 & 1115 $\pm$ 32 & 1107 $\pm$ 31 & $1091^{+27}_{-28}$\\
$f$& 1.6645 $\pm$ 0.0043 & 1.6643 $\pm$ 0.0045 & $1.661 \pm 0.019$ & $1.72 \pm 0.056$ & $1.660 \pm 0.011$ & $1.7187^{+0.0337}_{-0.0334}$ & 1.6645 $\pm$ 0.0107 \\
Age (Gyr) & 13.73 $\pm$ 0.12 & 14.08 $\pm$ 0.33 & $13.76^{+0.15}_{-0.14}$ & 14.49 $\pm$ 0.52 & 14.04 $\pm$ 0.36 & 14.48 $\pm$ 0.48 & $13.86^{+0.34}_{-0.33}$  \\
$\Omega_c h^2$ & 0.1139 $\pm$ 0.0041 & $0.1090^{+0.0060}_{-0.0061}$ & $0.1122^{+0.0068}_{-0.0069}$ & $0.1107^{+0.0063}_{-0.0062}$ & $0.1096^{+0.0061}_{-0.0062}$ & $0.1108^{+0.0060}_{-0.0061}$ & 0.1115 $\pm$ 0.0061  \\
$\Omega_{tot}$ &   - & $1.007^{+0.006}_{-0.007}$ &  - & 1.013 $\pm$ 0.007 & 1.006 $\pm$ 0.008 & 1.014 $\pm$ 0.007 & 1.003 $\pm$ 0.007  \\
$\sigma_8$ & 0.813 $\pm$ 0.028 & 0.787 $\pm$ 0.037 & $0.792^{+0.081}_{-0.082}$ & 0.907 $\pm$ 0.117 & $0.780^{+0.052}_{-0.053}$ & 0.904 $\pm$ 0.074 & $0.801^{+0.053}_{-0.052}$  \\
\hline
\end{tabular}
\caption{\label{table:baoowcdm} Marginalized one-dimensional
  constraints (68\%) for WMAP5$+$BAO for flat $\Lambda$CDM,
  $\Lambda$CDM with curvature (o$\Lambda$CDM), flat wCDM (wCDM), wCDM
  with curvature (owCDM), and owCDM including constraints from
  supernovae. The non-standard cosmological parameters constrained by
  the BAO measurements are $d_{0.275}\equiv r_s(z_d)/D_V(0.275)$ and
  $f\equiv D_V(0.35)/D_V(0.2)$.}
\end{center}
\end{table*}

We now turn to the constraints from our BAO measurement combined with
the full WMAP5 likelihood, including the constraint on $r_s(z_d)/D_A$ at
the time of decoupling.  While this extra constraint can break
degeneracies between $\Omega_m$, $\Omega_k$, and $w$ inherent in our
BAO constraints, the results are now sensitive to our assumption of a
constant dark energy equation of state $w$ at $z > 0.35$. Results for
the four models are presented in Table~\ref{table:baoowcdm}.

For the $\Lambda$CDM model, we find $\Omega_m = 0.278 \pm 0.018$ and
$H_0 = 70.1 \pm 1.5\Hunit$, with errors significantly reduced
compared to the WMAP5 alone analysis ($\Omega_m = 0.258 \pm 0.03$ and
$H_0 = 70.5^{+2.6}_{-2.7}\Hunit$). Similar limits on $\Omega_m$ were
obtained by \citet{rozo09} who used the maxBCG cluster abundance and
weak-lensing mass measurements to similarly break the tight WMAP5
constraint on $\Omega_mh^2$.
\begin{figure*}
  \centering
  \resizebox{0.7\textwidth}{!}{\includegraphics{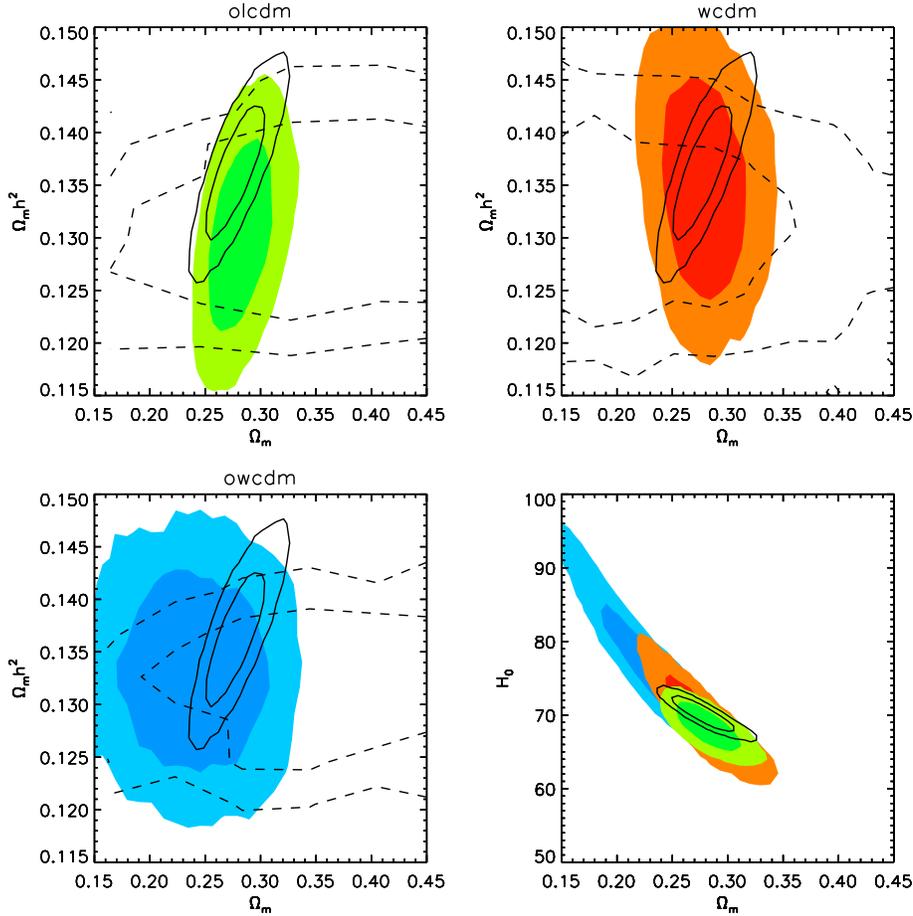}}
  \caption{\label{fig:baowmapgrid} WMAP5$+$BAO constraints on
    $\Omega_m h^2$, $\Omega_m$, and $H_0$ for $\Lambda$CDM (solid
    black contours), o$\Lambda$CDM (shaded green contours), wCDM
    (shaded red contours), and owCDM (shaded blue contours) models.
    Throughout, the solid contours show WMAP5$+$LRG $\Lambda$CDM
    constraints.  The first three panels show WMAP5 only constraints
    (dashed contours) and WMAP5$+$BAO constraints (colored contours)
    in the $\Omega_m h^2$- $\Omega_m$ plane as the model is varied.
    In the lower right we show all constraints from WMAP5$+$BAO for
    all four models in the $\Omega_m-H_0$ plane, which lie within the
    tight $\Omega_m h^2 \approx 0.133 \pm 0.006$ WMAP5-only
    constraints.}
\end{figure*}

Fig.~\ref{fig:baowmapgrid} shows the impact of relaxing the flat,
$\Lambda$CDM assumption.  The WMAP5 results alone tightly constrain
$\Omega_m h^2$ in all of these models (dashed lines), but low redshift
information is necessary to constrain $\Omega_m$ and $H_0$ separately.
Allowing $w \neq -1$ relaxes the constraint on $\Omega_m$ from the BAO
measurement, and in addition allowing $\Omega_k \neq 0$ relaxes the
constraint even further.  The impact on the constraints on $\Omega_m$
and $H_0$ is shown in the lower right panel.  All of the contours lie
along the banana with $\Omega_m h^2$ fixed from the CMB.

In the o$\Lambda$CDM model, the combination of scales measured by the
CMB and the BAO tightly constrain the curvature of the universe:
$\Omega_k = -0.007^{+0.006}_{-0.007}$.  The constraints on $\Omega_m$
and $H_0$ in this model are well described by
Eqns.~(\ref{eq:omconstraint}) \&~(\ref{eq:hconstraint}), while in the
wCDM cosmology they degrade because $w$ is not well-constrained by the
low redshift BAO information alone.

 \begin{figure}
  \centering
  \resizebox{0.9\columnwidth}{!}{\includegraphics{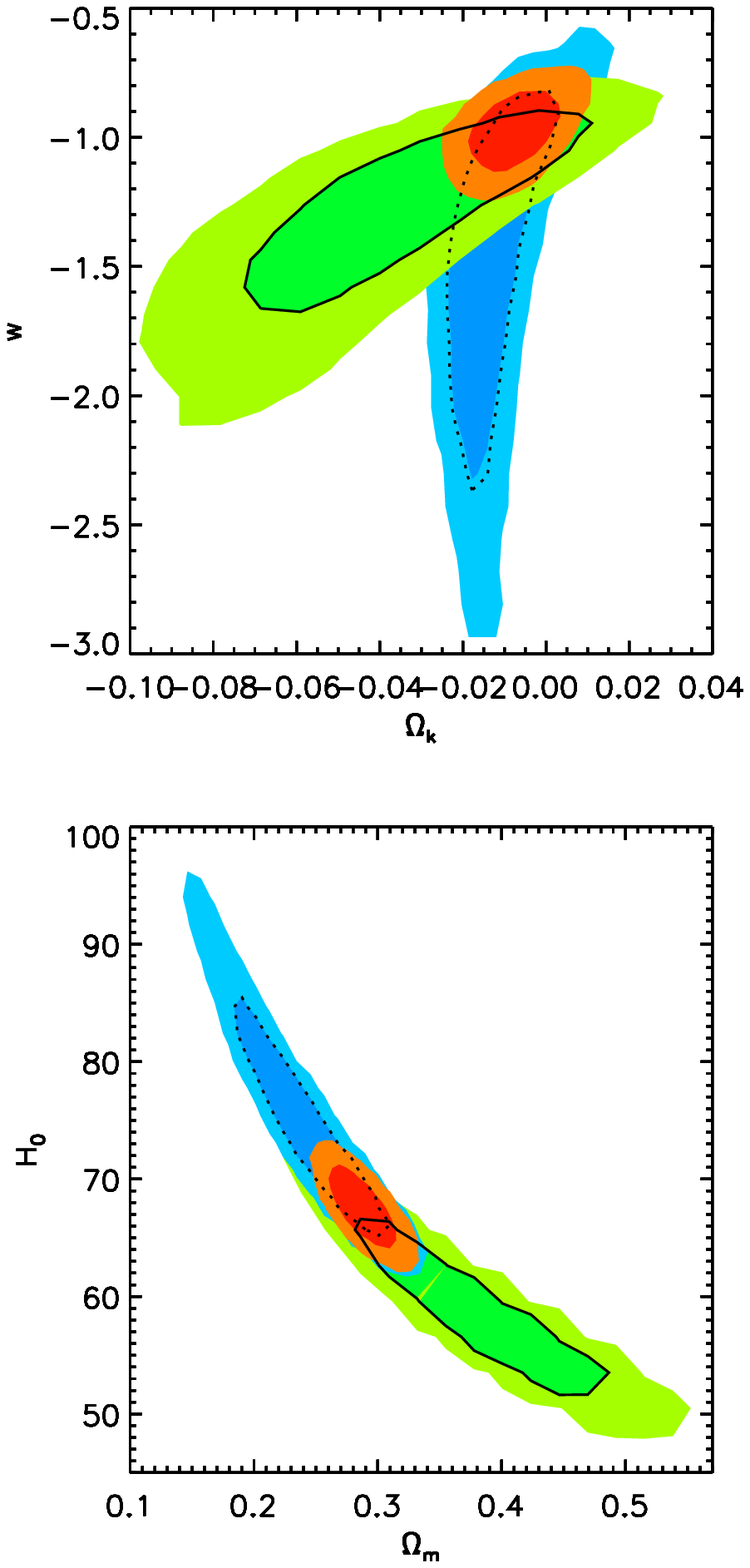}}
  \caption{\label{fig:omkwSN} For the owCDM model we compare the
    constraints from WMAP5$+$BAO (blue contours), WMAP5$+$SN (green
    contours), and WMAP5$+$BAO$+$SN (red contours). Dashed and solid
    contours highlight the 68\% confidence intervals for the
    WMAP5$+$BAO and WMAP5$+$SN models respectively.}
\end{figure}
When the parameter space is opened to both curvature and $w$, the
WMAP5 data are not able to eliminate the degeneracy between $\Omega_m$
and $w$ in the BAO constraint.  The constraints relax to $\Omega_m =
0.240^{+0.044}_{-0.043}$ and $H_0=75.3 \pm 7.1\Hunit$; $\Omega_k = -0.013
\pm 0.007$ is still well-constrained but $w$ is not (see
Fig.~\ref{fig:baowmapgrid}).  Including the constraints from the Union
Supernova Sample breaks the remaining degeneracy, and we recover the
tight constraints on $\Omega_m$ and $H_0$ given in
Eqns.~(\ref{eq:omconstraint}) \&~(\ref{eq:hconstraint}).  These
constraints, and the relative degeneracies induced and broken by
different data sets, are shown in Fig.~\ref{fig:omkwSN}. For each of
the four models considered, the central values for $\Omega_m$ and
$H_0$ change only slightly when the full WMAP5 likelihoods are used
(Table~\ref{table:baoowcdm}) instead of priors on $\Omega_b h^2$ and
$\Omega_c h^2$ in combination with the Union SN sample
(Table~\ref{table:baoSNandCMBprior}). 

Table~\ref{table:baoowcdm} also lists the best fit cosmological age
(i.e. time since the Big Bang) for different cosmologies and data sets.
While the age is very well determined for $\Lambda$CDM and wCDM, there
is a degeneracy between age and curvature that increases the
uncertainties and allows for an older age in o$\Lambda$CDM and owCDM
information.  Adding SN and $H_0$ measurements reduces these
uncertainties and implies a best fit age of $13.86_{-0.33}^{+0.34}$
Gyr.

\subsection{Comparison with \citet{riess/etal:2009} $H_0$}

\citet{riess/etal:2009} recently released a new determination of the
Hubble constant using a differential distance ladder: $H_0 = 74.2 \pm
3.6\Hunit$.  This value, as well as the values $H_0 \approx
68\Hunit$ determined in Table~\ref{table:baoSNandCMBprior} using BAO,
SN, and a WMAP5 prior on $\Omega_c h^2$ and $\Omega_b h^2$, are within
$\sim 1\sigma$ of the mean value determined from WMAP5+BAO in a
$\Lambda$CDM model, 70.1.  In the wCDM model, combining this new $H_0$
with the WMAP5 likelihood constrains $w = -1.12 \pm 0.12$.  In
Table~\ref{table:baoowcdm} we show MCMC results for the owCDM model
for WMAP5+BAO+$H_0$ and WMAP5+BAO+$H_0$+SN\footnote{We account for the
  small cosmology dependence in the $H_0$ constraint (seen as a slight
  degeneracy between $H_0$ and $w$ in fig. 14 of
  \citealt{riess/etal:2009}) by considering it as a constraint on the
  inverse luminosity distance at the effective redshift $z=0.04$
  (Riess private comm.)}.  In this model, the supernova data are more
effective than $H_0$ at breaking the long degeneracy in the WMAP5+BAO
constraints.  Combining WMAP5+BAO+SN+$H_0$, the mean parameters are
quite close to $\Lambda$CDM: $\Omega_k = -0.003 \pm 0.007$ and $w =
-1.00 \pm 0.10$, and $\Omega_m = 0.279 \pm 0.016$ and $H_0 = 69.5 \pm
2.0\Hunit$ are also well-constrained.

\section{Comparison with DR5 analyses}  \label{sec:cmpr_dr5}

In \citet{percival07c}, we presented BAO measurements calculated from
fitting power spectra calculated for three samples drawn from the
combined SDSS+2dFGRS catalogue, using the SDSS DR5 data. The full
catalogue was split into galaxy populations, rather than redshift
slices, corresponding to the SDSS LRGs, the 2dFGRS+SDSS main galaxies,
and the combined sample. From this, we obtained the distance
constraints $r_s(z_d)/D_V(0.2)=0.1980\pm0.0058$ and
$r_s(z_d)/D_V(0.35)=0.1094\pm0.0033$ with correlation coefficient
$0.39$, which gives a distance ratio measurement of
$D_V(0.35)/D_V(0.2)=1.812\pm0.062$. The concordance $\Lambda$CDM value
is $D_V(0.35)/D_V(0.2)=1.67$, measured using the SNLS supernova data,
which is discrepant with the published DR5 BAO results at the
2.4$\sigma$ level. The analysis of mock catalogues presented in
Section~\ref{sec:bao_ln} showed that the cubic spline $\times$ BAO
method underestimates the true distribution of recovered distances,
given noisy data, which produce a non-Gaussian likelihood surface. We
should therefore increase the errors on the DR5 measurements of
\citet{percival07c} by at least a factor of $1.14$, which is the
correction derived from the fits to three DR7 power spectra. If we do
this, the revised DR5 constraints are
$r_s(z_d)/D_V(0.2)=0.1981\pm0.0071$ and
$r_s(z_d)/D_V(0.35)=0.1094\pm0.0040$ with correlation coefficient
$0.38$, which gives a distance ratio measurement of
$D_V(0.35)/D_V(0.2)=1.813\pm0.073$. The discrepancy between the old
DR5 constraints and the SNLS $\Lambda$CDM value is reduced to $\simlt
2\sigma$. Because the DR5 data were noisier than the DR7 data, we
should expect the likelihood surface to be less like a Gaussian
prediction, and the correction actually should be slightly larger than
that for the DR7 data.

Of all the changes implemented between this DR7 analysis and the
analysis of the DR5 data, it was the increase in the number of random
points used to quantify the survey geometry that had the most effect
when comparing different catalogues. We now find consistent results,
given in Table~\ref{table:bao_DV_constraints}, for all catalogues and
analysis variations presented in Section~\ref{sec:robust}. When
translated into constraints on the distance ratio, for the full
catalogue we find $D_V(0.35)/D_V(0.2)=1.736\pm0.065$. Using only 3
redshift slices we find $D_V(0.35)/D_V(0.2)=1.765\pm0.079$. If the
0.5$\sigma$ difference is not due to chance, the difference between
these measurements could be caused by residual non-Gaussian scatter in
the band powers. A scenario in which this is reduced by including fits
to more redshift bins would then explain the observed trend. Excluding
the 2dFGRS and early SDSS data, the constraint is reduced to
$D_V(0.35)/D_V(0.2)=1.747\pm0.070$, which is consistent with the
tighter constraint using all of the data.

\citet{sanchez09}, who analysed the SDSS DR6 sample, speculated that
the discrepancy could be caused by the \citet{percival07c} analysis
fixing the BAO damping scale. However, in our current analysis, if we
allow the BAO damping scale $D_{\rm damp}$ to vary, the derived
constraints on $D_V(0.35)/D_V(0.2)$ does not change significantly from
that recovered in our default analysis. The mild discrepancy with
$\Lambda$CDM does not appear to be caused by fixing the damping
scale. The change from photometric calibration to uber-calibration
has a relatively minor effect on the distance ratio, which increases
to $D_V(0.35)/D_V(0.2)=1.748\pm0.074$. Fig.~\ref{fig:bao_cmpr_cc}
shows that the effect on the BAO of redshift-space distortions caused
by the thermal motion of galaxies in clusters is similarly
small. Linear redshift-space distortions propagate the apparent
position of galaxies along their velocity vector in a way that simply
makes the field look more evolved than it is; they do not alter the
positions of the BAO.

In conclusion, the significance of the discrepancy with flat
$\Lambda$CDM models is reduced because of
\begin{enumerate}
\item analysis of the non-Gaussian nature of the likelihood surface,
\item analysis of more redshift slices,
\item more accurate determination of the galaxy redshift distribution.
\end{enumerate}

\section{Discussion}  \label{sec:discuss}

In this paper we have measured and analysed BAO from the SDSS DR7
sample, which represents the final data set observed using the
original SDSS spectroscopic target selection algorithm. We have
further developed the analysis method used by \citet{percival07c} to
analyse the DR5 sample, including a faster method for the calculation
of the window function (see Appendix~\ref{app:win}), linking the
cosmological model to be tested with the power spectrum band-powers
measured. This has enabled us to analyse power spectra calculated for
six rather than three redshift slices, which would not have been
possible using the old method.

In Section~\ref{sec:results} we have shown how the distance--redshift
constraints at $z=0.2$ and $z=0.35$ can be decomposed into a single
distance constraint at $z=0.275$, and a ``gradient'' around this pivot
given by $D_V(0.35)/D_V(0.2)$. This allows us to easily test the
consistency of the $\Lambda$CDM model without having to compare with
additional data. For the best-fit flat $\Lambda$CDM model that matches
our constraint $d_{0.275} = 0.1390 \pm 0.0037$, we find that our
distance-ratio measurement of $D_V(0.35)/D_V(0.2)=1.736\pm0.065$ is
consistent at the 1.1$\sigma$ level.

Now that the SDSS-II sample is complete, the importance of including
the 2dFGRS data is reduced, and the inclusion only decreases the low
redshift $z=0.2$ distance error by 4\%. As we showed in
Section~\ref{sec:sample}, the inclusion of the 2dFGRS galaxies does
not lead to the discrepancy with the $\Lambda$CDM model: including the
2dFGRS brings our constraint slightly more into line with the
predictions of $\Lambda$CDM models.

Of the cosmological parameter constraints presented in
Tables~\ref{table:baoSNandCMBprior}, \&~\ref{table:baoowcdm}, perhaps
the most impressive are the constraints on $\Omega_m$ and $H_0$. For
$\Lambda$CDM models, fitting to BAO and Supernovae with priors on
$\Omega_mh^2$ and $\Omega_bh^2$ gives $H_0$ to 3.2\% and $\Omega_m$
to 6.4\%. These constraints are robust to the behaviour of the
Universe at high redshift, as they are based only on the
distance--redshift relation at redshift $z<0.35$: we can allow
$\Omega_k \neq 0$ and $w \neq -1$ with minimal effect. This weak
dependence on $w$ and $\Omega_k$ was shown in
Eqns.~(\ref{eq:omconstraint}) \&~(\ref{eq:hconstraint}) for the BAO
data.

If we allow for the flatness constraint to be relaxed, then we obtain
$\Omega_k=-0.007\pm0.007$ from the combination of BAO+WMAP5 data. A
tight constraint was similarly obtained on $w=-0.97\pm0.17$ if we
relax the $\Lambda$ constraint. If we allow both the curvature and the
dark energy equation of state to vary, we must include more data to
continue to break the degeneracy between the two parameters. We do so
by including results from the Union SN dataset, giving us $\Omega_k =
-0.006\pm0.008$ and $w=-0.97 \pm 0.10$, consistent with a flat
$\Lambda$CDM model. If one allows only $w\ne-1$ OR $\Omega_k\ne0$,
then the combination of CMB, supernova and BAO data has an internal
cross-check: opening two degrees of freedom from flat $\Lambda$CDM
yields results that are consistent with flat $\Lambda$CDM. We have
also shown that our constraints are consistent with the recent
re-determination of $H_0$ by \citet{riess/etal:2009}, and that
combining this constraint with WMAP5, BAO, and SN in a model where
both curvature and $w$ vary yields mean parameter values very close to
$\Lambda$CDM.

In a companion paper \citep{reid09}, we consider the LRG sample in
more detail. The LRGs are distributed in haloes in a simple way and we
are able to extract the halo power spectrum from the data. In addition
to fitting the BAO in this power spectrum, we are able to extract
limited information about the shape of the power, which gives
complementary constraints. A detailed comparison between the results
from our fit to the BAO in redshift slices, performed in a cosmology
model-independent way and including low-redshift galaxies, with the
halo power spectrum of \citet{reid09} is presented in that paper,
where excellent agreement is demonstrated. The data sets are
correlated so they should not be used together to constrain
cosmological models.

Our analysis highlights the importance of BAO as a key method for
investigating cosmic acceleration, and shows that the method can
already provide interesting cosmological constraints. Ongoing
spectroscopic surveys aiming to use BAO to analyse dark energy include
the Baryon Oscillation Spectroscopic Survay (BOSS;
\citealt{schlegel09a}), the Hobby-Eberly Dark Energy Experiment
(HETDEX; \citealt{hill08}) and the WiggleZ survey
\citep{glazebrook07}. There are also plans for future surveys covering
significantly larger volumes of the Universe, and therefore observing
the BAO signal with higher precision such as the Square Kilometer
Array (SKA: {\tt www.skatelescope.org}), and the Joint Dark Energy
Mission (JDEM: {\tt jdem.gsfc.nasa.gov}) and European Space Agency
Euclid satellite mission concepts, or the Big Baryon Oscillation
Spectroscopic Survay (BigBOSS; \citealt{schlegel09b}). Photometric
surveys such as the Dark Energy Survey (DES: {\tt
  www.darkenergysurvey.org}), the Panoramic Survey Telescope \& Rapid
Response System (Pan-Starrs: {\tt pan-starrs.ifa.hawaii.edu}) and the
Large Synoptic Survey Telescope (LSST: {\tt www.lsst.org}) will find
BAO using photometric redshifts. All of these surveys will measure BAO
at higher redshifts than those analysed in our paper using SDSS-II
data: if dark energy does not have a simple explanation, then
comparison between future high redshift results and our current
understanding of the low-redshift Universe from SDSS-II will provide
an interesting test of these models.

\section*{Acknowledgements}

WJP is grateful for support from the UK Science and Technology
Facilities Council, the Leverhulme trust and the European Research
Council.  DJE was supported by National Science Foundation grant
AST-0707225 and NASA grant NNX07AC51G.  Simulated catalogues were
calculated and analysed using the COSMOS Altix 3700 supercomputer, a
UK-CCC facility supported by HEFCE and STFC in cooperation with
CGI/Intel.  WJP would like to thank Tamara M. Davis, Ravi K. Sheth,
Roman Scoccimarro, Eyal Kazin, Taka Matsubara and the referee, Fergus
Simpson, for useful interactions.

The 2dF Galaxy Redshift Survey was undertaken using the Two-degree Field
facility on the 3.9m Anglo-Australian Telescope. The success of the survey
was made possible by the dedicated efforts of the staff of the
Anglo-Australian Observatory, both in creating the 2dF instrument and
in supporting the survey observations.

Funding for the SDSS and SDSS-II has been provided by the Alfred
P. Sloan Foundation, the Participating Institutions, the National
Science Foundation, the U.S. Department of Energy, the National
Aeronautics and Space Administration, the Japanese Monbukagakusho, the
Max Planck Society, and the Higher Education Funding Council for
England. The SDSS Web Site is {\tt http://www.sdss.org/}.

The SDSS is managed by the Astrophysical Research Consortium for the
Participating Institutions. The Participating Institutions are the
American Museum of Natural History, Astrophysical Institute Potsdam,
University of Basel, Cambridge University, Case Western Reserve
University, University of Chicago, Drexel University, Fermilab, the
Institute for Advanced Study, the Japan Participation Group, Johns
Hopkins University, the Joint Institute for Nuclear Astrophysics, the
Kavli Institute for Particle Astrophysics and Cosmology, the Korean
Scientist Group, the Chinese Academy of Sciences (LAMOST), Los Alamos
National Laboratory, the Max-Planck-Institute for Astronomy (MPIA),
the Max-Planck-Institute for Astrophysics (MPA), New Mexico State
University, Ohio State University, University of Pittsburgh,
University of Portsmouth, Princeton University, the United States
Naval Observatory, and the University of Washington.

\setlength{\bibhang}{2.0em}

\appendix

\section{Calculation of the window function} \label{app:win}

In this Appendix, we describe the method used to calculate the mapping
between the power spectra in the ``true'' cosmology to be tested, and
the measured, or observed, power spectra where a $\Lambda$CDM model
was used to convert redshifts to distances. This window function
includes both the effect of the survey geometry and the mapping
between cosmological models. As described by \citet{percival07c}, we
should expect the observed power spectrum to be a convolution of the
true power spectrum with a window function.
\begin{equation}  \label{eq:Wint}
  P(k)_{\rm obs}=\int\,dk'W(k,k')P(k')_{\rm true}.
\end{equation}
The goal of this section is to introduce a fast method by which
$W(k,k')$ can be calculated for any model.

In \citet{percival07c}, this window function was calculated using
Monte-Carlo realisations of Gaussian density fields, created assuming
the cosmological model to be tested. These fields were then distorted
as if they had been analysed assuming a $\Lambda$CDM model, and the
power spectrum was calculated and compared with that input. Using a
large number of simple input power spectra, we were able to construct
the window function from this comparison. This procedure required
significant computational resources as many density fields were needed
in order to accurately measure the window function, limiting the
number of models that could be tested. In particular, we were only
able to consider cubic spline models of $D_V(z)$ with two nodes to
three power spectra. With a faster window function calculation, we can
include more nodes, and fit to more power spectra.

For a survey covering a thin shell, the window function relating true
and observed power is an offset delta function
\begin{equation}  \label{eq:W_narrow}
  W(k,k')=\delta_D[k/k'-\epsilon],
\end{equation} 
where $\epsilon=d_p({\rm true})/d_p({\rm obs})$ is the ratio of proper
distances in the true and observed cosmologies. Here we are simply
stretching the true survey prior to measuring the power spectrum.

\begin{figure}
  \centering
  \resizebox{0.9\columnwidth}{!}{\includegraphics{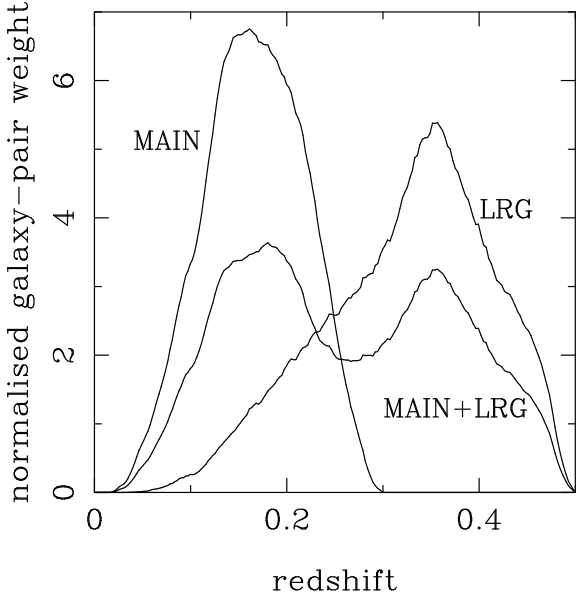}}
  \caption{The redshift dependence of galaxy pair-weights for the SDSS
    DR7 LRG and main galaxy samples, and from the combination of the
    two. These curves were calculated assuming a flat $\Lambda$CDM
    cosmology with $\Omega_m=0.25$, $h=0.72$, \& $\Omega_bh^2=0.0223$.}
  \label{fig:pair_weight}
\end{figure}
The obvious extension to surveys over a range of redshifts is to split
the sample into $i$ redshift shells, and to approximate the window
function as
\begin{equation}  \label{eq:W_broad}
  W(k,k')=\sum_i \delta_D[k/k'-\epsilon_i]w_i,
\end{equation}
where $w_i$ is the weighted number of galaxy pairs in redshift shell
$i$. Because we are now considering a broad survey, this pair weight
is a function of pair separation. In this paper, we bin pairs of
galaxies with comoving separation $90\mpcoh<d_{\Lambda{\rm
    CDM}}<130\mpcoh$, where $d_{\Lambda{\rm CDM}}$ is the comoving
distance in the $\Lambda$CDM cosmology used to convert galaxy
redshifts to distances. The bin size was chosen to approximately match
the BAO scale.  For the SDSS LRG, main galaxy and combined samples,
the galaxy pair-weights are shown in Fig.~\ref{fig:pair_weight}. We
also need to allow for differences in the orientation of galaxy pairs,
as the distribution of $\epsilon_i$ should allow the galaxy pairs to
be of all orientations. Including radial separations introduces an
asymmetric convolution for $\epsilon_i$, and we have found that this
needs to be included in order to provide approximately the correct
window function shapes.  Note that Eq.~(\ref{eq:W_broad}) is exact
when there is a perfect dilation of scale between the true and
observed cosmologies: such stretching of the windows can be perfectly
represented by this equation.

\begin{figure}
  \centering
  \resizebox{0.9\columnwidth}{!}{\includegraphics{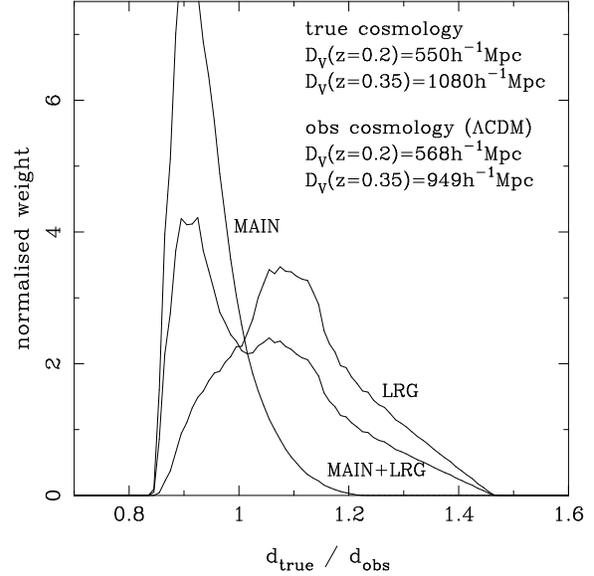}}
  \caption{Galaxy pair-weights for the SDSS DR7 LRG and main galaxy
    samples, and from the combination of the two, as a function of
    comoving distance shifts. These were calculated assuming that a
    flat $\Lambda$CDM cosmology with $\Omega_m=0.25$, $h=0.72$, \&
    $\Omega_bh^2=0.0223$ was used to analyse the data, while the BAO
    are present in a true cosmological model with distance--redshift
    relation defined by a cubic spline in $D_V(z)$ with nodes at
    $z=0.2$ and $z=0.35$, with amplitude as shown in the plot.}
  \label{fig:dratio}
\end{figure}
For each ``true'' cosmology to be tested, we can calculate the shift
in scale that stretches each pair of galaxies because we do not
measure BAO using this model. We have to allow for the angular shift
caused by a change in $D_A(z)$ and the radial shift caused by the true
and observed $H(z)$ being different. An example of the weighted
distribution of ``shifts'' expected for a model cosmology defined by a
cubic spline in $D_V(z)$ with two nodes at $z=0.2$ and $z=0.35$ is
shown in Fig.~\ref{fig:dratio}. Here the true cosmology has a
distance-redshift relation given by a spline fit to $D_V(z)$, with
nodes $D_V(z=0.2)=550\mpcoh$, and $D_V(z=0.35)=1080\mpcoh$. The
$\Lambda$CDM values are $D_V(z=0.2)=568\mpcoh$, and
$D_V(z=0.35)=949\mpcoh$, so at redshift $z=0.2$, BAO in the true
cosmology are stretched to larger scales by the analysis method, while
those at redshift $z=0.35$ are compressed to smaller scales. For the
SDSS main galaxies, with median redshift close to $z\simeq0.2$,
$d_{\rm true}/d_{\rm obs}<1$, while for the LRGs, with median redshift
$z\simeq0.35$, $d_{\rm true}/d_{\rm obs}>1$.

\begin{figure*}
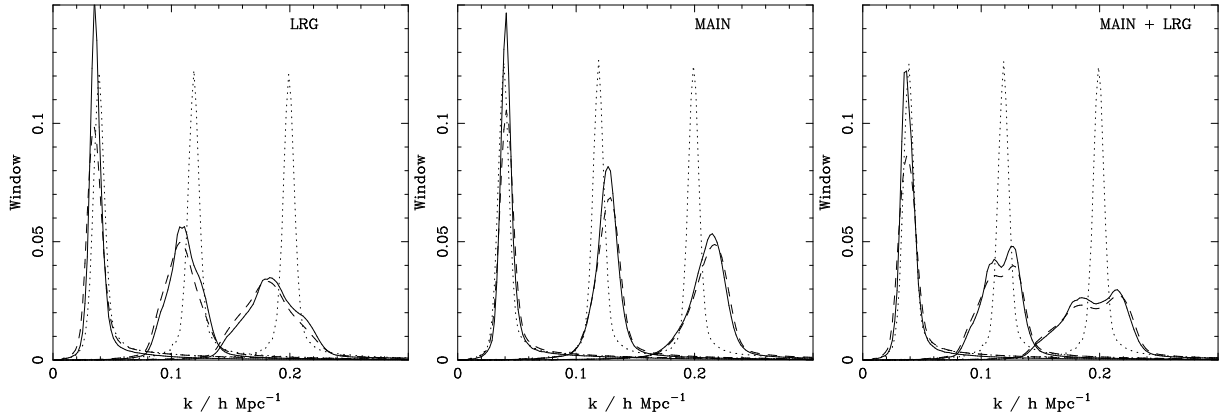

  \centering
  \resizebox{0.3\textwidth}{!}{\includegraphics{win_cmpr_LRG.ps}}
  \resizebox{0.3\textwidth}{!}{\includegraphics{win_cmpr_MAIN.ps}}
  \resizebox{0.3\textwidth}{!}{\includegraphics{win_cmpr_COMB.ps}}
  \caption{Window functions for three values of $k$, calculated for
    the SDSS LRG, main galaxy and combined catalogues. Dotted lines
    represent the windows for our fiducial $\Lambda$CDM
    cosmology. Solid and dashed lines show the window functions, if
    the true cosmology were different, but the data were analysed
    assuming that the fiducial $\Lambda$CDM cosmology is correct. The
    solid lines were calculated using the procedure outlined in this
    Appendix. Dashed lines were calculated using the Monte-Carlo
    procedure of \citet{percival07c}.}
  \label{fig:win_cmpr}
\end{figure*}
For each ``true'' cosmological model, the window function relating the
true and observed power spectra was calculated by convolving the
standard window function for the $\Lambda$CDM model, by the
distribution of shifts such as that shown in
Fig.~\ref{fig:dratio}. For the models shown in Fig.~\ref{fig:dratio},
we have calculated the window function using the approximate method
outlined in this Appendix, and using the Monte-Carlo method described
by \citet{percival07c}. A comparison of the windows is presented in
Fig.~\ref{fig:win_cmpr}. Reasonable agreement is found between the
different methods: it is clear that the approximate method of
splitting into shells recovers the main features of the window
function. The agreement is not perfect, as expected given the
approximate nature of our calculation. Because we analyse the data
using a $\Lambda$CDM model, the window will be correct for this model,
and will only deviate if we consider significantly different
distance--redshift relations.

\label{lastpage}

\end{document}